\documentclass[aps,prd,nofootinbib]{revtex4}
\usepackage[utf8]{inputenc}
\usepackage[english]{babel}
\usepackage[T1]{fontenc}

\usepackage{amsmath}
\usepackage{amsfonts}
\usepackage{amssymb}
\usepackage{graphicx}

\usepackage{lmodern}

\usepackage{hyperref}
\hypersetup{colorlinks, 
citecolor=blue,
linkcolor=red,
urlcolor=black}

\usepackage[normalem]{ulem}

\usepackage[title]{appendix}

\usepackage{booktabs}
\usepackage{multirow}

\begin{document}

%\title{Induced Quantum Brownian motion by different boundary conditions in Minkowski spacetime}
%\title{Induced Brownian motion in inhomogeneous tridimentional space and $S^{1}\times R^{3}$ modified space-time.}
%\title{Quantum Brownian motion induced by an inhomogeneous tridimensional space and a topologically $S^{1}\times R^{3}$ modified space-time.}
\title{Quantum Brownian motion induced by an inhomogeneous tridimensional space and a $S^{1}\times R^{3}$ topological space-time}
\author{Éwerton J. B. Ferreira}
\email{ejbf@academico.ufpb.br}
\author{Eliza M. B. Guedes}
\email{eliza.brito@academico.ufpb.br}
\author{Herondy F. Santana Mota}
\email{hmota@fisica.ufpb.br}
%\author{author3}
%\email{autor3@email.com}
\affiliation{Departamento de Física, Universidade Federal da Paraíba, Caixa Postal 5008, João Pessoa, Paraíba, Brazil}

%\date{\today}

\begin{abstract}
In this paper we investigate the Quantum Brownian motion of a point particle induced by quantum vacuum fluctuations of a massless scalar field in $(3+1)$-dimensional Minkowski spacetime with distinct conditions (Dirichlet, Neumann, mixed and quasiperiodic). The modes of the field are confined and compactified to a finite length region, which consequently provides a natural measure scale for the system. Useful expressions for the Wightman function have been obtained, which allow us to calculate analytical expressions for the velocity dispersion in all condition cases considered. We also obtain expressions for the velocity dispersion in the short and late time regimes. Finally, we exhibit some graphs in order to show the behavior of the velocity dispersions, discussing important divergencies that are present in our results. 
\end{abstract}

%\pacs{87.01-a}

\maketitle

\section{Introduction} %%%%%%%%%%%%%%%%%%%%%%%%%%%%%%%%%%%%%%%%%%%%%%%%%%%%%%%%%%%%%%%%%%%%%%%%%%%%%%

	The stochastic motion performed by a point particle when interacting with the quantum vacuum fluctuations of a relativistic field, e.g., scalar or electromagnetic, is also known as Quantum Brownian motion (QBM). This is an example of a phenomena class which arise from quantum vacuum fluctuations and that, over the past several years, has been studied in different scenarios and with different approaches  \cite{gour1999will,yu2004vacuum,yu2004brownian,yu2006brownian,seriu2008switching,seriu2009smearing,
bessa2009brownian,de2016probing,de2019remarks,de2014quantum,camargo2018vacuum,camargo2019vacuum,Camargo:2020fxp,mota2020induced,anacleto2021stochastic,ferreira2022quantum, bessa2017quantum}. The quantum vacuum fluctuations are always present but only become observable when the vacuum is somehow perturbed, for instance, by considering elements such as boundary conditions, temperature, nontrivial topology and so on. 

Similarly to the classical Brownian motion, the typical quantities for the quantum version that should be investigated are the position and velocity dispersions. However, the analogy between the classical and quantum Brownian motion is limited given that in the quantum scenario the dispersions can assume negative values, something that does not occur  in the classical case. In the latter, dispersions are quantities positively defined, that is, $\langle(\Delta A)^{2}\rangle > 0$, where $A$ is some physical observable to be measured. So, negative values for the dispersions in the classical scenario does not make sense. On the other hand, in the quantum context it is possible that $\langle(\Delta A)^{2}\rangle < 0$, which can be interpreted as due to quantum uncertainty reduction \cite{yu2004brownian,yu2004vacuum}, subvacuum effects \cite{camargo2018vacuum,Camargo:2020fxp, camargo2019vacuum} and failure in the renormalization process as a consequence of boundary conditions imposed on the field \cite{de2014quantum}.	

In the quantum context, the basic idea is that a point particle (structureless) interacting with quantum vacuum fluctuations of a field has an induced stochastic motion. In the electromagnetic case, for instance, it has been analyzed in Refs. \cite{yu2004vacuum} and \cite{yu2004brownian} the QBM as a consequence of one and two perfectly reflecting parallel planes, respectively. In both cases the position and velocity dispersions are calculated. Moreover, the study of thermal effects for the QBM in the electromagnetic case with one perfectly reflecting plane was also developed in Ref. {\cite{yu2006brownian}, where the magnitude of thermal and quantum contributions are discussed. Thereby, it is shown that for well defined temperature regimes one contribution can be more significant than the other. By seeking to investigate more realistic systems, a wave packet like structure for the particle has been proposed in Ref. \cite{seriu2009smearing}. In addition, switching time effects associated with the interaction between a point particle and the quantum vacuum fluctuations of the field are considered in Refs.  \cite{seriu2008switching,de2016probing}. Also, switching time effects at finite temperature are taken into account in Ref. \cite{de2019remarks}.
	
Regarding the QBM induced by vacuum fluctuations of a quantum scalar field, the investigations conducted follow similarly to the electromagnetic case. In Ref. \cite{de2014quantum}, for instance, it is studied the induced QBM due to a massless scalar field in the presence of a perfectly reflecting plane and in Ref. \cite{camargo2018vacuum} switching time effects are taken into account. The massive scalar field case in $(D+1)$-dimensions, with one perfectly reflecting plane, is studied in Ref. \cite{camargo2019vacuum} and thermal effects in Ref. \cite{Camargo:2020fxp}, both also considering switching time effects. All scalar field cases just mentioned use Dirichlet boundary condition, which set a null value for the field modes on the boundary. As a complement to the study of the QBM, it is worth mentioning that it has also been investigated in the cosmological context, in particular, considering dark matter detection. According to this scenario, in principle, dark matter may induce a stochastic motion in a test particle of ordinary matter, whose observation would offer new insights into the understanding of dark matter properties \cite{Cheng:2019vwy}.

As a contribution to all cases considered in literature so far, for the massless scalar field, we intend to take into consideration two elements, as far as we know, not yet explored in the study of the QBM. The first of them is to consider, in analogy to the electromagnetic case, two perfectly reflecting parallel planes where the scalar field satisfies not only Dirichlet but also Neumann and mixed boundary conditions (BC's). This way, we confine the modes of the field in one direction, something that naturally leads to momentum discretization in the same direction, providing a natural scale for the system. The second element we would like to consider is the effect of a quasiperiodic condition on the QBM of a scalar point particle. 

The conditions mentioned in the previous paragraph can also be seen as possible ways of alter the topology of the spatial section of the Minkowski spacetime, which is the background where we are performing our investigations. The consideration of two planes, for instance, breaks the homogeneity and isotropy of space, which can be interpreted as a way of simulating a topological modification in the spatial section of the spacetime. In the case of the induced inhomogeneity, we note that, in the presence of planes, the spatial directions $y$ and $z$ are similar, but differ from the $x$ direction, where the planes are located. In fact, an observer in the $yz$ plane will perceive an infinity bidimensional space, but the same observer on the $xy$ or $xz$ planes will perceive a semi-infinity space, that is, infinity in the $y$ and $z$ directions, but finite in the $x$ direction. On the other hand, the anisotropy, as we shall see, it is shown by the distinction between the velocity dispersions, which is the observable investigated in this work.

It is important to mention that the investigation of the induced Brownian motion considering nontrivial topologies for the spatial section of the Minkowski spacetime is a topic that has been explored for the past several years. Recently, in Ref. \cite{Bessa:2019aar}, it was investigated the Brownian motion of a point particle induced by quantum vacum fluctuations of an electromagnetic field in a flat spacetime whose spatial section has nontrivial topologies. In principle, it is suggested that this effect can be used to indicate the global inhomogeneity of space. 
For similar and more recent discussions see also Refs. \cite{Lemos:2020ogj}, \cite{Lemos:2021rya} and \cite{Lemos:2022rms}, where these effects as a function of their time evolution are used as a supposed indicator of spatial orientability. See also Ref. \cite{Lemos:2021jzy} for an example in a conformally expanding flat spacetime.
In addition, we would like to point out that, in this context, the Casimir effect has also been investigated; for more details see for instance Ref. \cite{Sutter:2006dj} and references therein. Therefore, taking into consideration the current status of the subject just described, in the case of the massless scalar field, the present work aims to complement the investigations conducted so far for the QBM.

It is also worth to emphasize that BC's are not merely technical and mathematical details of academic interest, they also can be related to physical properties of the studied systems. Dirichlet and Neumann BC's, for instance, specify the field value and its normal derivate on the boundary, respectively. Typically, we found these conditions in electrostatic systems where either an electrical potential is fixed on the surface (Dirichlet BC) or the corresponding electric field ($\nabla\phi$) is the one fixed on the surface (Neumann BC) \cite{jackson1998classical, arfken2005mathematical}. Also, there exists mixed BC's, in which case both the field and its normal derivative are specified on the boundary. In Ref. \cite{alves2000spontaneous}, for instance, the spontaneous emission of a two-level system between two parallel plates has been investigated taking into consideration that the electromagnetic vector potential obeys BC's similar to the mixed one on the plates. In this case, one of the plates is perfectly conducting and the other one is perfectly permeable. Hence, we can say that mixed BC's simulates plates with distinct physical properties. As to the quasiperiodic condition, it can indicate the existence of an interaction present in the system. As an example, we can mention the Aharonov-Bohm effect \cite{de2012topological, kretzschmar1965must}.
	
	Regarding the structure of this paper, in Section \ref{SecWF} we describe the system to be investigated, indicating some useful simplifications. Then, we exhibit the complete set of normalized solutions for the scalar field for each condition used in this work. This allows us to obtain the positive frequency Wightman function for each case, which is a fundamental element in our calculations. We also obtain a  general form for the Wightman function representing Dirichlet, Neumann and mixed BC's in a single expression. In Section \ref{SecVelDis} we calculate the particle velocity dispersion. Finally, in Section \ref{Conclusions}, we present our conclusions summarizing the main results obtained. Note that we have also dedicated Appendices \ref{AppA} and \ref{AppB} to obtain important expressions used to investigate asymptotic limits for the velocity dispersions. In this work we use natural units such that $c=\hbar=1$. 

%%%%%%%%%%%%%%%%%%%%%%%%%%%%%%%%%%%%%%
\section{Wightman Functions}\label{SecWF}
%%%%%%%%%%%%%%%%%%%%%%%%%%%%%%%%%%%%%%
\subsection{Model, general field solution and the expression to calculate the Wightman function}
%%%%%%%%%%%%%%%%%%%%%%%%%%%%%%%%%%%%%%
%
In this section we want to establish some important results that will be used later on in the velocity dispersion computation, namely, the complete set of normalized solutions for the scalar field and the corresponding Wightman functions. In other words, we are interested in investigating the induced QBM of a point particle coupled to a fluctuating quantum massless scalar field, considering different conditions. As it is known, this stochastic motion is induced by the quantum vacuum fluctuations of the field. The classical action that describes this system is written as
%and which we can quantize to obtain the necessary expressions

\begin{eqnarray}\label{eq01}
S_{\text{tot}} = S_{\text{f}} + S_{\text{p}} + S_{\text{int}},
\end{eqnarray}
where
\begin{eqnarray}
S_{\text{f}} &=& \int dt\int dV \dfrac{(\partial_{\mu}\phi)(\partial^{\mu}\phi)}{2},
\end{eqnarray}
is the massless scalar field part of the action,
\begin{eqnarray}
S_{\text{p}} &=& \int dt \dfrac{m{\bf\dot{ x}}^{2}}{2} 
\end{eqnarray}
corresponds to the action describing a point particle of mass $m$ and
\begin{eqnarray}
S_{\text{int}} &=&-g\int dt\int dV\delta^3({\bf x}-{\bf x'})\phi
\end{eqnarray}
stands for the interaction between the particle and the massless scalar field $\phi$, $dV$ is the volume element of the spatial section of the spacetime and $\delta^3({\bf x}-{\bf x'})$ is the spatial three dimensional Dirac delta function. Note that the measure of the strength of the interaction, denoted by $g$, is the charge of the point particle. This is a model widely known in the literature and has been considered in different scenarios \cite{gour1999will,de2014quantum,camargo2018vacuum,de2019remarks,camargo2019vacuum,Camargo:2020fxp}.

The variation of the action \eqref{eq01} with respect to field, $\phi$, provides the massless Klein-Gordon equation with a three dimensional Dirac delta function as a source, that is,
\begin{eqnarray}\label{eq02}
\Box\phi({\bf x},t) = -g\delta^3({\bf x}-{\bf x'}),
\end{eqnarray}
where $\Box =\partial_{\nu}\partial^{\mu}$ is the d'Alembertian differential operator to be considered in Minkowski spacetime described by the line element
\begin{equation}
ds^2 = dt^2 - dx^2 - dy^2 - dz^2.
\end{equation}
Thereby, Eq. \eqref{eq02} is the equation of motion for a massless scalar field coupled to a point particle with mass $m$ and charge $g$. Although this is a non-homogeneous differential equation, we wish to consider that the point particle's influence on the field is negligible \cite{de2014quantum}. This allows us to write Eq. \eqref{eq02} as
\begin{eqnarray}
\Box\phi({\bf x},t) \approx 0,
\end{eqnarray}
which gives a general nonnormalized solution in terms of plane waves, i.e.,
\begin{eqnarray}\label{eq03}
\phi_{\sigma}({\bf x},t)=Ne^{-i\omega t + ik_{x}x +ik_{y}y+ ik_{z}z},
\end{eqnarray}
where $\omega^{2} = k_{x}^{2}+k_{y}^{2}+k_{z}^{2}$ are the eigenfrequencies of the field, with $k_i$ being the momentum in each spatial direction, and $\sigma = (k_x, k_y, k_z)$ stands for the set of quantum numbers.
The constant $N$ can be obtained via normalization condition
\begin{eqnarray}\label{eq04}
2\omega\int dV\phi_{\sigma}(w)\phi_{\sigma '}^{*}(w)=\delta_{\sigma\sigma '},
\end{eqnarray}
where the delta symbol in the r.h.s is understood as Kronecker delta for discrete quantum numbers and Dirac delta function for continuous quantum numbers. Note that we have introduced the notation $w=({\bf x},t)$ to specify spacetime coordinates. As we shall see later, the solution in Eq. \eqref{eq03} is modified when subjected to both the boundary conditions and the quasiperiodic condition, leading to discretization of one of the momenta.

Once subjecting the solution in Eq. \eqref{eq03} to the conditions considered, we must find the normalization constant $N$ by making use of Eq. \eqref{eq04}. This process makes possible to write the complete set of normalized solution and use it to calculate the Wightman function, which is a crucial element to our computations. In order to construct the Wightman function, we may first promote the field to an operator and write it in terms of the positive and negative frequency normalized solutions, with coefficients of the expansion being the creation $a_{\sigma}^{\dagger}$ and annihilation $a_{\sigma}$ operators. Mathematically, we use the standard construction \cite{birrell1984quantum}
\begin{eqnarray}\label{eq05qf}
\hat{\phi}(w) = \sum_{\sigma}[a_{\sigma}\phi_{\sigma}(w)+a_{\sigma}^{\dagger}\phi_{\sigma}^{*}(w)],
\end{eqnarray}
where the creation and annihilation operators obey the commutation relation $[a_{\sigma},a_{\sigma'}^{\dagger}]=\delta_{\sigma\sigma'} $. We, thus, are able to obtain the Wightman function by taking into consideration the definition
\begin{eqnarray}\label{eq05wf}
W(w,w')=\langle 0|\hat{\phi}(w)\hat{\phi}(w')|0\rangle = \sum_{\sigma}\phi_{\sigma}(w)\phi_{\sigma}^{*}(w'),
\end{eqnarray}
where $|0\rangle$ is the vacuum state of the scalar field. Hence, the above equation provides the positive frequency Wightman function for the scalar field. In addition, the summation symbol in \eqref{eq05wf} stands for either integrals in the continuous quantum numbers or possible sums over discrete ones.

%%%%%%%%%%%%%%%%%%%%%%%%%%%%%%%%%
\subsection{Dirichlet boundary condition}
%%%%%%%%%%%%%%%%%%%%%%%%%%%%%%%%%
Firstly we are interested in considering Dirichlet boundary condition on the massless scalar field solution \eqref{eq03}. This means that by confining the field in a region of length $a$ between two perfectly reflecting parallel planes, perpendicular to the $x$-direction, we must have the condition 
\begin{eqnarray}\label{eq06}
\phi({\bf x},t)|_{x=0}=\phi({\bf x},t)|_{x=a}=0.
\end{eqnarray}
Therefore, from Eqs. \eqref{eq03}, \eqref{eq06} and \eqref{eq04} we find that the complete set of normalized solutions in this case is given by
\begin{eqnarray}
\phi_{\sigma}({\bf x},t) = \dfrac{1}{\sqrt{4\pi^{2}a\omega_{n}}}\sin(k_{n}x)e^{-i\omega_{n}t+ik_{y}y+ik_{z}z},
\end{eqnarray}
where $\omega_{n}^2=k_{n}^{2}+k_{y}^{2}+k_{z}^2$ are the eigenfrequencies of the field, with momentum in the $x$-direction now discretized, that is, $k_{n}=\frac{n\pi}{a}$ $(n=1,2,3,\ldots)$. The set of quantum numbers in this case is $\sigma=(n, k_y, k_z)$. In Fig.\ref{figtwoparallelplanesA} we give an illustration of the setup described above. This configuration will also be used for the cases of Neumann and mixed boundary conditions later on.
\begin{figure}[h]
\includegraphics[scale=0.2]{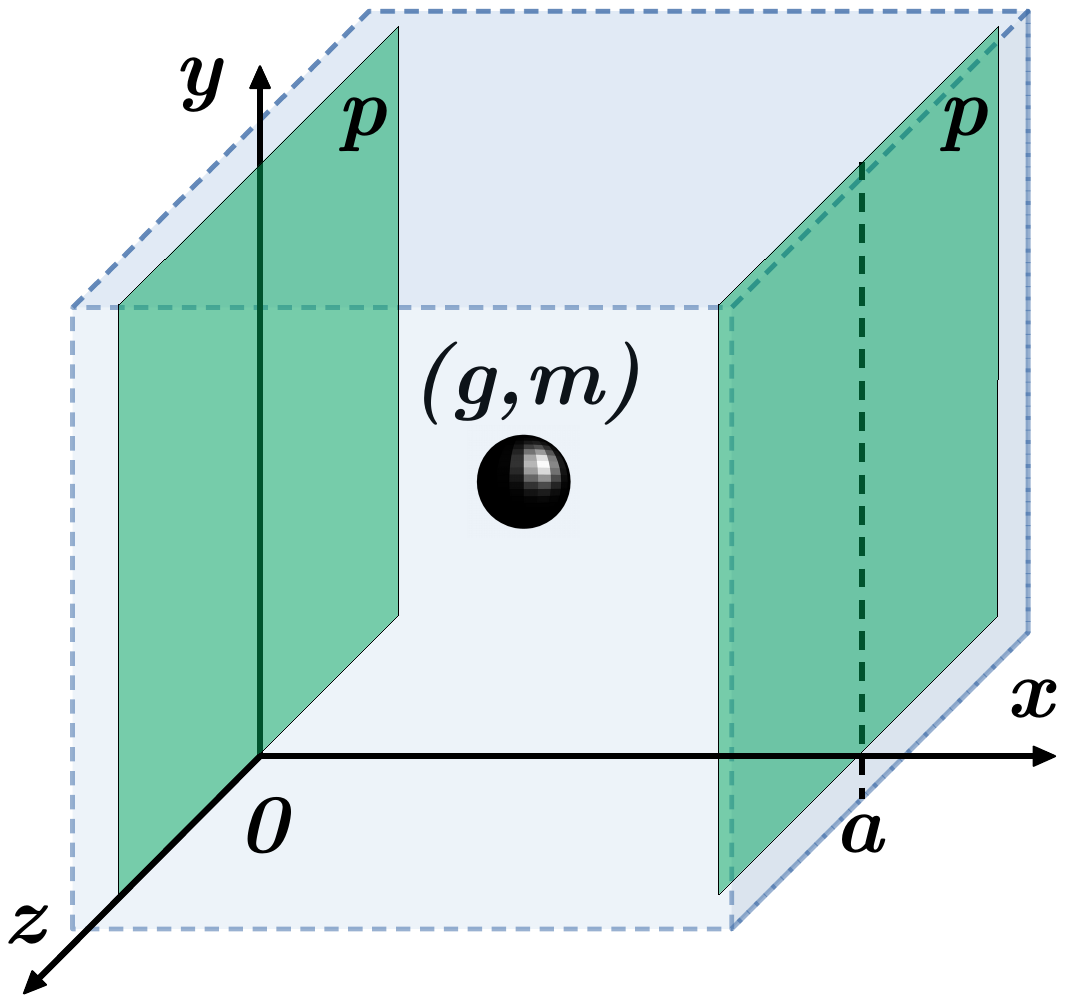}
\caption{A point particle with mass $m$ and charge $g$ in the presence of two identical and perfectly reflecting parallel planes $p$ placed at $x=0$ and $x=a$, confining the field modes of a massless quantum scalar field.}\label{figtwoparallelplanesA}
\end{figure}

In order to obtain the corresponding Wightman function we make use of Eq. \eqref{eq05wf} with the summation symbol defined as
\begin{equation}
 \sum_{\sigma}\equiv\sum_{n=1}^{\infty}\int_{-\infty}^{\infty}dk_{y}\int_{-\infty}^{\infty}dk_{z}.
 \label{SSD}
\end{equation}
Consequently, the Wightman function becomes
\begin{eqnarray}\label{eq07}
W^{\text{(D)}} = \dfrac{1}{2\pi a}\sum_{n=1}^{\infty}\int_{0}^{\infty}dk k J_{0}(\Delta\ell k)\sin(k_{n}x)\sin(k_{n}x')\dfrac{e^{-i\omega_{n} \Delta t}}{\omega_{n}},
\end{eqnarray}
where $J_{\mu}(z)$ is the Bessel function \cite{gradshtein2007}, $\Delta\ell = \sqrt{\Delta y^{2}+\Delta z^{2}}$, $\Delta y = y-y'$, $\Delta z = z-z'$ and $\Delta t = t-t'$. Note that in the above expression we have used polar coordinates for the plane defined by the momentum variables $k_{y}$ and $k_{z}$, such that $k^{2}=k_{y}^{2}+k_{z}^{2}$ and $dk_{y}dk_{z}\rightarrow kdkd\theta$, which made possible to perform the angular integral leading to the Bessel function.

The sum in $n$ present in the Wightman function expression in Eq. \eqref{eq07} can be worked out by making use of the Abel-Plana formula \cite{saharian2007generalized}
\begin{eqnarray}\label{eqAbelPlana}
\sum_{n=0}^{\infty}F(n)=\dfrac{1}{2}F(0) + \int_{0}^{\infty}d\xi F(\xi) + i\int_{0}^{\infty}d\xi\dfrac{[F(i\xi)-F(-i\xi)]}{e^{2\pi \xi}-1}.
\end{eqnarray}
This is a very useful expression and it is often used, for example, in the Casimir energy computations (see Ref. \cite{saharian2007generalized} for more details). The function $F(n)$ in the present case is taken to be
\begin{eqnarray}
F(n) = \sin(k_{n}x)\sin(k_{n}x')\dfrac{e^{-i\omega_{n} \Delta t}}{\omega_{n}},
\label{funF}
\end{eqnarray} 
where $F(0)=0$ and, consequently, the contribution from the first term in the r.h.s. of \eqref{eqAbelPlana} vanishes. Hence, by using the above expression in Eq. \eqref{eqAbelPlana}, after some algebraic manipulations, Eq. \eqref{eq07} can be written as 
\begin{eqnarray}
W^{\text{(D)}}=W_{1}^{\text{(D)}}+W_{2}^{\text{(D)}},
\label{WFDcomp}
\end{eqnarray} 
where for mathematical clarity and convenience, after the change of variables $s=\frac{\pi\xi}{a}$, we have defined
\begin{eqnarray}\label{eqW1dirichlet}
W_{1}^{\text{(D)}} =\dfrac{1}{2\pi{^2}}\int_{0}^{\infty}ds\sin(sx)\sin(sx')\int_{0}^{\infty}dk\dfrac{kJ_{0}(\Delta\ell k)e^{-i\Delta t\sqrt{k^{2}+s^{2}}}}{\sqrt{k^{2}+s^{2}}},
\end{eqnarray}
and
\begin{eqnarray}\label{eqW2dirichlet}
W_{2}^{\text{(D)}}=\dfrac{1}{\pi^{2}}\int_{0}^{\infty}dk k J_{0}(\Delta\ell k)\int_{k}^{\infty}ds\dfrac{\sin(isx)\sin(isx')}{e^{2as}-1}\dfrac{\cosh(\Delta t\sqrt{s^{2}-k^{2}})}{\sqrt{s^{2}-k^{2}}}.
\end{eqnarray}
Note that the expression in Eq. \eqref{eqW1dirichlet} stems from the integral in the second term in the r.h.s of the Abel-Plana formula \eqref{eqAbelPlana}, while Eq. \eqref{eqW2dirichlet} stems from the third term. In the latter, we have also used the identity
\begin{equation}
\sqrt{(\pm i s)^2 + k^2}=\left\{ \begin{array}{l}\pm i\sqrt{s^2 - k^2},\qquad\mathrm{for}\,\, s>k\,,\\
\;\;\;\sqrt{k^2 - s^2},\,\,\qquad\mathrm{for}\,\, s<k\,.
\end{array}\right.
\label{identity}
\end{equation}
The integrals in Eqs. \eqref{eqW1dirichlet} and \eqref{eqW2dirichlet} can be solved with the help of Refs. \cite{prudnikov1986integrals,gradshtein2007}, providing the expressions
\begin{eqnarray}\label{eqW1dirichletsolved}
W_{1}^{\text{(D)}} = \frac{1}{4\pi^2}\left\{\frac{1}{[\Delta x^2 + \alpha^2]} - \frac{1}{[\Delta\bar{x}^2 + \alpha^2]}\right\},
\end{eqnarray}
and
\begin{eqnarray}\label{eqW2dirichletsolved}
W_{2}^{\text{(D)}}=-W_{1}^{\text{(D)}} + \dfrac{1}{8\pi a\alpha}\left\{ \dfrac{\sinh\left(\frac{\pi\alpha}{a}\right)}{\left[  \cosh\left(\frac{\pi\alpha}{a}\right)-\cos\left(\frac{\pi\Delta x}{a}\right) \right]}-\dfrac{\sinh\left(\frac{\pi\alpha}{a}\right)}{\left[  \cosh\left(\frac{\pi\alpha}{a}\right)-\cos\left(\frac{\pi\Delta \bar{x}}{a}\right) \right]}  \right\}.
\end{eqnarray}
Consequently, by substituting the two results above in Eq. \eqref{WFDcomp} we obtain
\begin{eqnarray}\label{eqW3dirichlet}
W^{\text{(D)}}=\dfrac{1}{8\pi a\alpha}\left\{ \dfrac{\sinh\left(\frac{\pi\alpha}{a}\right)}{\left[  \cosh\left(\frac{\pi\alpha}{a}\right)-\cos\left(\frac{\pi\Delta x}{a}\right) \right]}-\dfrac{\sinh\left(\frac{\pi\alpha}{a}\right)}{\left[  \cosh\left(\frac{\pi\alpha}{a}\right)-\cos\left(\frac{\pi\Delta \bar{x}}{a}\right) \right]}  \right\},
\end{eqnarray}
where $\alpha^{2}=\Delta y^{2}+\Delta z^{2}-\Delta t^{2}$, with $\Delta x = x-x'$ and $\Delta\bar{x} = x+x'$. For our purposes, we can further simplify Eq. \eqref{eqW3dirichlet} by using the identity \cite{gradshtein2007}
\begin{eqnarray}\label{eq08}
\dfrac{\sinh\left(\frac{\pi\alpha}{a}\right)}{\left[  \cosh\left(\frac{\pi\alpha}{a}\right)-\cos\left(\frac{\pi\Delta x}{a}\right) \right]} = \dfrac{2a\alpha}{\pi}\sum_{n=-\infty}^{\infty}\dfrac{1}{[(\Delta x - 2an)^{2}+\alpha^{2}]}.
\end{eqnarray}
This is particularly useful since we can separate the Minkowski contribution in a clearer way.  This contribution is the term $n=0$ of the sum, which is divergent in the coincidence limit $w'\rightarrow w$. As it is known, this divergent contribution must be subtracted from the calculation of the velocity dispersion in order to obtain a renormalized quantity. Therefore, Eq. \eqref{eqW3dirichlet} takes the form
\begin{eqnarray}\label{eq09WFDirichlet}
W^{\text{(D)}}(w,w')=\dfrac{1}{4\pi^2}\sum_{n=-\infty}^{\infty}\left[f_{n}(\Delta r)-f_{n}(\Delta \bar{r})\right],
\end{eqnarray}
where 
\begin{eqnarray}\label{eq09fn}
f_{n}(\Delta r) &=& \dfrac{1}{(\Delta x -2an)^2+\Delta y^2+\Delta z^2 - \Delta t^2},\nonumber\\
f_{n}(\Delta \bar{r}) &=& \dfrac{1}{(\Delta \bar{x} -2an)^2+\Delta y^2+\Delta z^2 - \Delta t^2}.
\end{eqnarray}
Hence, Eq. \eqref{eq09WFDirichlet} correspond to the positive frequency Wightman function in cartesian coordinates for the massless scalar field whose modes are restrict to obey Dirichlet boundary condition on the two perfectly reflecting parallel planes, placed at $x=0$ and $x=a$. Note that the Minkowski contribution comes from the term $n=0$ in the function $f_{n}(\Delta r) $. In contrast, the term $n=0$ is finite in the coincidence limit $w'\rightarrow w$ for the function $f_{n}(\Delta \bar{r})$. It is in fact the one plane contribution of the Wightman function for the Dirichlet boundary condition.

%
%%%%%%%%%%%%%%%%%%%%%%%%%%%
\subsection{Neumann boundary condition}
%%%%%%%%%%%%%%%%%%%%%%%%%%%
%
In the case we use Neumann boundary condition, the normal derivative of the field must vanish in the boundary. In this sense, considering two perfectly reflecting parallel planes, placed at $x=0$ and $x=a$, we have
\begin{eqnarray}\label{eq10}
\left[\partial_{x}\phi({\bf x},t)\right]|_{x=0}=\left[\partial_{x}\phi({\bf x},t)\right]|_{x=a}=0.
\end{eqnarray}
So, from Eqs. \eqref{eq03}, \eqref{eq10} and \eqref{eq04} we obtain the complete set of normalized solutions as follows 
\begin{eqnarray}
\phi({\bf x},t) = c_{n}\cos(k_{n}x)e^{-i\omega_{n}t+ik_{y}y+ik_{z}z},
\end{eqnarray}
where the eigenfrequencies are given by $\omega_{n}^2=k_{n}^{2}+k_{y}^{2}+k_{z}^2$, with $k_{n}=\frac{n\pi}{a}$  $(n=0,1,2,\ldots)$ being the discretized momentum in the $x$-direction, and $\sigma = (n, k_y, k_z)$ is the set of quantum numbers. The normalization constant $c_n$ is written as
\begin{eqnarray}
c_{n} = \left\{
	\begin{matrix}
		\dfrac{1}{\sqrt{8\pi^{2}a\omega_{n}}}, & n = 0, \\
		\dfrac{1}{\sqrt{4\pi^{2}a\omega_{n}}}, & n\geq 1.
	\end{matrix}\right.
\end{eqnarray}

Similarly to the previous case, we can calculate the Wightman function by making use of Eq. \eqref{eq05wf}. The summation symbol in Eq. \eqref{eq05wf} is now defined as 
\begin{equation}
 \sum_{\sigma}\equiv\sum_{n=0}^{\infty}\int_{-\infty}^{\infty}dk_{y}\int_{-\infty}^{\infty}dk_{z}.
 \label{SSD2}
\end{equation}
Consequently, the Wightman function takes the form
\begin{eqnarray}\label{eq11}
W^{\text{(N)}}&=& W_1^{\text{(N)}} + W^{\text{(D)}}\nonumber\\
&=&\dfrac{1}{4\pi^{2}a}\sum_{n=0}^{\infty}c^{*}_{n}\int_{-\infty}^{\infty}dk_{y}\int_{-\infty}^{\infty}dk_{z}\dfrac{\cos(k_{n}\Delta\bar{x})e^{-i\omega_{n}\Delta t+ik_{y}\Delta y+ik_{z}\Delta z}}{\omega_{n}} + W^{(D)},
\end{eqnarray}
where $c^{*}_{0}=1/2$ and $c^{*}_{n\geq 1}=1$. In the above expression, we have used trigonometric identities in order to be possible to identify two contributions, i.e., the one in the first term in the r.h.s and the one in the second term corresponding to the Wightman function for Dirichlet boundary condition, given by Eq. \eqref{eq09WFDirichlet}. Since the Dirichlet part has previously been calculated, we only need to focus in the first term in the r.h.s. of Eq. \eqref{eq11}. In the end, the Wightman function for the Neumann boundary condition case takes into consideration the sum of both terms in Eq. \eqref{eq11}. 

Let us then work out the first term in the r.h.s. of Eq. \eqref{eq11}. This is possible with the help of the identity 
\begin{eqnarray}
\dfrac{e^{-\omega_{n}\Delta\tau}}{\omega_{n}}=\dfrac{2}{\sqrt{\pi}}\int_{0}^{\infty}ds e^{-\omega_{n}^{2}s^{2}-\frac{\Delta\tau^{2}}{4s^{2}}},
\end{eqnarray}
where we have performed the Wick rotation, $\Delta\tau=i\Delta t$. The use of the above identity in $ W_1^{(N)} $, along with the help of Ref. \cite{{gradshtein2007}}, leads to
\begin{eqnarray}\label{eq12}
W^{\text{(N)}}_{1}&=&\dfrac{1}{2\pi a\alpha}\sum_{n=0}^{\infty}c_{n}^{*}\cos\left(\dfrac{\pi\Delta\bar{x}}{a}\right)e^{-\left(\frac{\pi\alpha}{a}\right)n}\nonumber\\
&=&\dfrac{1}{4\pi a\alpha}\dfrac{\sinh\left(\frac{\pi\alpha}{a}\right)}{\left[  \cosh\left(\frac{\pi\alpha}{a}\right)-\cos\left(\frac{\pi\Delta \bar{x}}{a}\right) \right]}.
\end{eqnarray}
Hence, in view of Eqs. \eqref{eq11}, \eqref{eq12} and \eqref{eqW3dirichlet}, we conclude that
\begin{eqnarray}\label{eq09WFNeumann}
W^{\text{(N)}}(w,w')=\dfrac{1}{4\pi^2}\sum_{n=-\infty}^{\infty}\left[f_{n}(\Delta r)+f_{n}(\Delta \bar{r})\right],
\end{eqnarray}
where the functions defined in \eqref{eq09fn} have been used. This is the positive frequency Wightman function for the massless scalar field obeying Neumann boundary condition on the two perfectly reflecting parallel planes. Note 
that the difference between the expressions for the Wightman function in the Dirichlet and Neumann boundary condition cases consists of only a changing of sign in the second term of Eq. \eqref{eq09WFDirichlet}. Again, the contribution $n=0$ of the sum in the first term in the r.h.s. of Eq. \eqref{eq09WFNeumann} is the divergent Minkowski contribution in the coincidence limit $w'\rightarrow w$, while in the second term is the one plane finite contribution. The latter has the opposite sign when compared to the Dirichlet boundary condition case.

%
%%%%%%%%%%%%%%%%%%%%%%%%%%%%%%%%%%%%%%%
\subsection{Mixed boundary condition}
%%%%%%%%%%%%%%%%%%%%%%%%%%%%%%%%%%%%%%%
%
In the mixed boundary condition case, the general solution of the field in Eq. \eqref{eq03} must obey Dirichlet condition in one plane and Neumann condition in the other. Thus, two configurations are possible on the first and second planes, that is, Dirichlet and Neumann (DN) as well as Neumann and Dirichlet (ND). For the configuration DN, respectively at $x=0$ and $x=a$, the condition obeyed by the field is given by 
\begin{eqnarray}\label{eq13}
\phi({\bf x},t)|_{x=0}=\left[\partial_{x}\phi({\bf x},t)\right]|_{x=a}=0.
\end{eqnarray}
By applying the condition \eqref{eq13} on Eq. \eqref{eq03}, with the use of Eq. \eqref{eq04} afterwards, we obtain the complete set of normalized solutions
\begin{eqnarray}\label{eq14}
\phi_{\sigma}({\bf x},t) = \dfrac{1}{\sqrt{4\pi^{2}a\omega_{n}}}\sin(k_{n}x)e^{-i\omega_{n}t+ik_{y}y+ik_{z}z},
\end{eqnarray}
where the eigenfrequencies are now written as $\omega_{n}^2=k_{n}^{2}+k_{y}^{2}+k_{z}^2$, with $k_{n}=\frac{\pi(2n+1)}{2a}$ $(n=0,1,2, \ldots)$. Again, the momentum in the $x$-direction has been discretized as a consequence of Eq. \eqref{eq13} and the set of quantum numbers is specified by $\sigma = (n, k_y, k_z)$.

The Wightman function is computed through Eq. \eqref{eq05wf}, by making use of the normalized solution in Eq. \eqref{eq14} and
\begin{equation}
 \sum_{\sigma}\equiv\sum_{n=0}^{\infty}\int_{-\infty}^{\infty}dk_{y}\int_{-\infty}^{\infty}dk_{z}.
 \label{SSD3}
\end{equation}
Thereby, similarly to the Dirichlet condition case, it is possible to write the Wightman function in the form
\begin{eqnarray}\label{eq15}
W^{\text{(M)}} = \dfrac{1}{2\pi a}\sum_{n=0}^{\infty}\int_{0}^{\infty}dk k J_{0}(\Delta\ell k)\sin(k_{n}x)\sin(k_{n}x')\dfrac{e^{-i\omega_{n} \Delta t}}{\omega_{n}},
\end{eqnarray}
where we have used again polar coordinates for the plane defined by the momentum variables $k_y$ and $k_z$, following the same steps as in the Dirichlet condition case. By taking into account the structure of the allowed values for $k_{n}$ it is more convenient to use the Abel-Plana formula written in the form \cite{saharian2007generalized}
\begin{eqnarray}\label{eqAbelPlana2}
\sum_{n=0}^{\infty}F\left(n+\frac{1}{2}\right)= \int_{0}^{\infty}d\xi F(\xi) - i\int_{0}^{\infty}d\xi\dfrac{[F(i\xi)-F(-i\xi)]}{e^{2\pi \xi}+1},
\end{eqnarray} 
where the function $F\left(n+\frac{1}{2}\right)$ is defined as in Eq. \eqref{funF} but now with $k_{n}=\frac{\pi(2n+1)}{2a}$. The Abel-Plana formula above allows us to write the Wightman function as 
\begin{eqnarray}
W^{\text{(M)}}&=&W^{\text{(D)}}_{1} + W^{\text{(M)}}_{1}\nonumber\\
&=&W^{\text{(D)}}_{1} - \dfrac{1}{\pi^{2}}\int_{0}^{\infty}dk k J_{0}(\Delta\ell k)\int_{k}^{\infty}ds\dfrac{\sin(isx)\sin(isx')}{e^{2as}+1}\dfrac{\cosh(\Delta t\sqrt{s^{2}-k^{2}})}{\sqrt{s^{2}-k^{2}}} ,
\label{WFM}
\end{eqnarray}
where $W^{\text{(D)}}_{1}$ is given by Eq. \eqref{eqW1dirichletsolved} and we have again made use of the identity \eqref{identity}. Furthermore, the contribution in the second term in the r.h.s. of the above expression is found to be 
\begin{eqnarray}
W^{\text{(M)}}_{1} = -W_{1}^{\text{(D)}} + \dfrac{1}{4\pi a\alpha}\left\{ \dfrac{\sinh\left(\frac{\pi\alpha}{2a}\right)\cos\left(\frac{\Delta x\pi}{2a}\right)}{\left[  \cosh\left(\frac{\pi\alpha}{a}\right)-\cos\left(\frac{\pi\Delta x}{a}\right) \right]}-\dfrac{\sinh\left(\frac{\pi\alpha}{2a}\right)\cos\left(\frac{\Delta \bar{x}\pi}{2a}\right)}{\left[  \cosh\left(\frac{\pi\alpha}{a}\right)-\cos\left(\frac{\pi\Delta \bar{x}}{a}\right) \right]}  \right\},
\label{firstC}
\end{eqnarray}
where we have also again used the help of Refs. \cite{gradshtein2007, prudnikov1986integrals} to solve the integrals in $k$ and in $s$.

The complete Wightman function for the mixed boundary condition case is obtained from Eq. \eqref{WFM}, by using the expressions in Eqs. \eqref{eqW1dirichletsolved} and \eqref{firstC}. This gives
\begin{eqnarray}\label{eqWmistas}
W^{\text{(M)}}=\dfrac{1}{4\pi a\alpha}\left\{ \dfrac{\sinh\left(\frac{\pi\alpha}{2a}\right)\cos\left(\frac{\pi\Delta x}{2a}\right)}{\left[  \cosh\left(\frac{\pi\alpha}{a}\right)-\cos\left(\frac{\pi\Delta x}{a}\right) \right]}-\dfrac{\sinh\left(\frac{\pi\alpha}{2a}\right)\cos\left(\frac{\pi\Delta \bar{x}}{2a}\right)}{\left[  \cosh\left(\frac{\pi\alpha}{a}\right)-\cos\left(\frac{\pi\Delta \bar{x}}{a}\right) \right]}  \right\}.
\end{eqnarray} 
We can still put the above expression in a more convenient form, analogously to what has been done to the Dirichlet and Neumann condition cases. So, let us make use of the identity \cite{gradshtein2007}%
\begin{eqnarray}
\dfrac{\sinh\left(\frac{\pi\alpha}{2a}\right)\cos\left(\frac{\pi\Delta x}{2a}\right)}{\left[  \cosh\left(\frac{\pi\alpha}{a}\right)-\cos\left(\frac{\pi\Delta x}{a}\right) \right]} = \dfrac{\alpha a}{\pi}\sum_{n=-\infty}^{\infty}\dfrac{e^{i\pi n}}{[(\Delta x - 2an)^{2}+\alpha^{2}]}.
\end{eqnarray} 
Consequently,
\begin{eqnarray}\label{eq16WFMistas}
W^{\text{(M)}}(w,w')=\dfrac{1}{4\pi^2}\sum_{n=-\infty}^{\infty}(-1)^{n}\left[f_{n}(\Delta r)-f_{n}(\Delta \bar{r})\right],
\end{eqnarray} 
where the functions introduced in \eqref{eq09fn} have been used. It is important to observe that in Eq. \eqref{eq13} we use a configuration of boundary conditions such that Dirichlet and Neumann conditions are applied to the planes at $x=0$ and $x=a$, respectively. If the reverse configuration is used,  that is, Neumann and Dirichlet such that $(\partial_{x}\phi)|_{x=0}=\phi|_{x=a}=0$, proceeding in a similar way as above, we obtain the same result shown in Eq. \eqref{eq16WFMistas}, but with the opposite sign in the second term in the r.h.s., which becomes positive. 
%This sign comes from the fact that when we use the configuration ND arise a cosine functions in the integrand of the Eq. \eqref{eq16}. This change in the second term signal is similar to that occur when pure Dirichlet or Neumann boundary conditions are used for both plates.

The results in Eqs. \eqref{eq09WFDirichlet}, \eqref{eq09WFNeumann} and \eqref{eq16WFMistas} obtained for the Wightman function in the cases of Dirichlet, Neumann and mixed boundary conditions can be written as a general and compact expression, i.e.,
\begin{eqnarray}\label{FWDNM}
W^{\text{(i)}}(w,w')=\dfrac{1}{4\pi^{2}} \sum_{n=-\infty}^{\infty}\left[ \gamma_{n}^{\text{(i)}}f_{n}(\Delta r)+\delta_{n}^{\text{(i)}}f_{n}(\Delta \bar{r}) \right],
\end{eqnarray}
where we have conveniently defined 
\begin{eqnarray}\label{Coefgd}
\gamma_{n}^{\text{(i)}}&=&\left[\gamma_{n}^{\text{(D)}},\gamma_{n}^{\text{(N)}},\gamma_{n}^{\text{(DN)}},\gamma_{n}^{\text{(ND)}}\right]=[+1,+1,(-1)^{n},(-1)^{n}],\nonumber \\
\delta_{n}^{\text{(i)}}&=&\left[\delta_{n}^{\text{(D)}},\delta_{n}^{\text{(N)}},\delta_{n}^{\text{(DN)}},\delta_{n}^{\text{(ND)}}\right]=[-1,+1,(-1)^{n+1},(-1)^{n}].
\end{eqnarray}
We can note that for all three boundary condition cases analyzed so far the contribution $n=0$ in the sum present in the first term in the r.h.s. of Eq. \eqref{FWDNM} correspond to the Minkowski contribution, which as we have already remarked, is divergent in the coincidence limit $w'\rightarrow w$. This term, as usual, must be subtracted from the physical observables. Moreover, the contribution $n=0$ coming from the second term in the r.h.s. of Eq. \eqref{FWDNM} provides the known expression for only one plane, placed at position $x=0$. The way we have organized the obtained Wightman functions in only one compact expression in Eq. \eqref{FWDNM} is very useful in the sense that it allows us to calculate at once the velocity dispersion for all three boundary condition cases since the derivative and integration operations in Eq. \eqref{eq21}, necessary to calculate the velocity dispersion, will only affect the functions $f_{n}$. Hence, after solving the successive operations acting on $f_{n}$ to obtain the velocity dispersion, we may just select the appropriate coefficients $\gamma_{n}^{(i)}$ and $\delta_{n}^{(i)}$ in order to specify which boundary condition result we are interested in.

%%%%%%%%%%%%%%%%%%%%%%%%%%%%%%%%
\subsection{Quasiperiodic condition}
%%%%%%%%%%%%%%%%%%%%%%%%%%%%%%%%
%
Finally, we now wish to consider a quasiperiodic condition, which generalizes the well known periodic and antiperiodic conditions by introducing a constant phase $\beta$, that is,
\begin{eqnarray}\label{eq17}
\phi(x,y,z,t) = e^{-2\pi\beta i}\phi(x+a,y,z,t).
\end{eqnarray}
The quasiperiodic parameter $\beta$ assumes values in the range $0\leq\beta <1$. Note that, if $\beta=0$ we restore the periodic condition whereas if $\beta=1/2$ we recover the antiperiodic one. Hence, the boundary condition above allows us to obtain a solution for the scalar field which includes besides the well known periodic and antiperiodic condition particular cases, also the cases for which $\beta\neq 0, 1/2$. As it is clear from Eq. \eqref{eq17}, we consider that the compactification, of length $a$, is in the $x$-direction. An illustrative representation of this four-dimensional spacetime configuration is shown in Fig.\ref{figR3xS1}. The introduction of the quasiperiodic parameter $\beta$ may be thought of representing possible interactions in the system, as in the case of the well known Aharonov-Bohm effect \cite{de2012topological, kretzschmar1965must}.

\begin{figure}[h]
\includegraphics[scale=0.3]{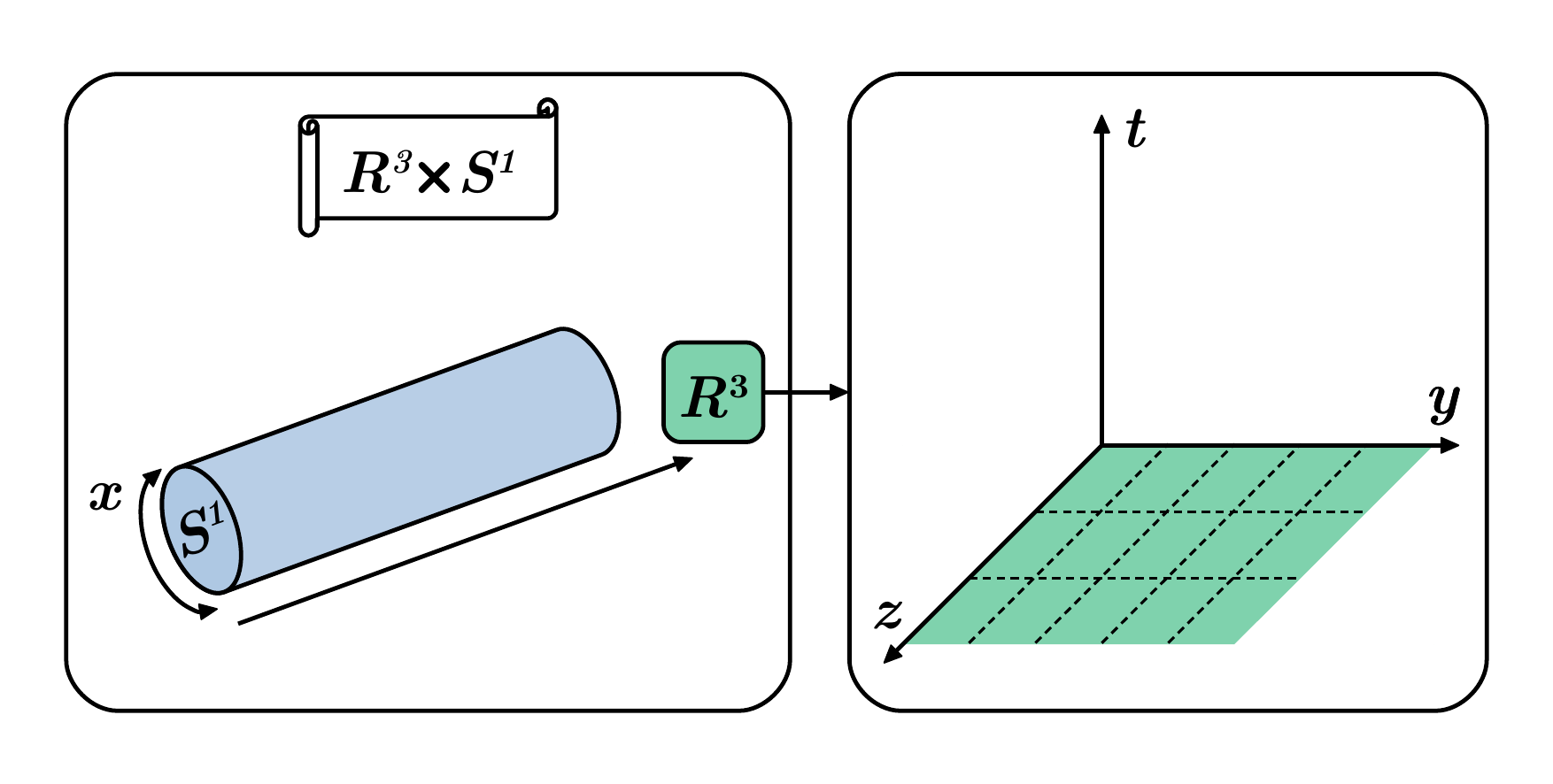}
\caption{Illustrative representation of four-dimensional spacetime with a compactified spatial dimension. The spacetime is composed of a compactified spatial dimension $x$, $S_{1}$, and the tridimensional space $R^{3}$ of coordinates $t,y,z$.}\label{figR3xS1}
\end{figure}

By requiring the solution in Eq. \eqref{eq03} to obey the condition \eqref{eq17}, after making use of the normalization condition \eqref{eq04}, we find 
\begin{eqnarray}
\phi_{\sigma}({\bf x},t) = \dfrac{1}{\sqrt{8\pi^{2}a\omega_{n}}}e^{-i\omega_{n}t+ik_{n}x+ik_{y}y+ik_{z}z},
\end{eqnarray}
where the eigenfrequencies are written as $\omega_{n}^2=k_{n}^{2}+k_{y}^{2}+k_{z}^2$, $k_{n} = \frac{2\pi(n+\beta)}{a}$ $(n=0,\pm 1,\pm 2,\ldots)$ and the set of quantum numbers is represented by $\sigma = (n, k_y, k_z)$. Similarly to the previous computations, one is able to calculate the Wightman function through Eq.  \eqref{eq05wf}, with
\begin{equation}
 \sum_{\sigma}\equiv\sum_{n=-\infty}^{\infty}\int_{-\infty}^{\infty}dk_{y}\int_{-\infty}^{\infty}dk_{z}.
 \label{SSD3QP}
\end{equation}
Next, we again adopt polar coordinates in the $(k_{y},k_{z})$-plane, such that $dk_{y}dk_{z}\rightarrow kdkd\theta$ and $k^{2}=k_{y}^{2}+k_{z}^{2}$. After solving the angular part we found
\begin{eqnarray}
W(w,w')&=&\dfrac{1}{4\pi a}\sum_{n=-\infty}^{\infty}e^{ik_{n}\Delta x}\int_{0}^{\infty}dk \dfrac{kJ_{0}(\Delta r k)e^{-i\omega_{n}\Delta t}}{\omega_{n}}\nonumber\\
&=&\dfrac{1}{4\pi a}\sum_{n=-\infty}^{\infty}e^{ik_{n}\Delta x-\alpha |k_{n}|},
\label{WFQPn}
\end{eqnarray}
    where we have used the help of Ref. \cite{{prudnikov1986integrals}} to solve the integral in $k$. By splitting the summation in $n$ in two parts in order to eliminate the modulus in $k_n$, with the help of Ref. \cite{gradshtein2007}, we can further simplify the expression in the second line of Eq. \eqref{WFQPn} and write it in the convenient form
\begin{eqnarray}\label{eq18WFQuasiperiodicas}
W(w,w')=\dfrac{1}{4\pi^{2}}\sum_{n=-\infty}^{\infty}e^{2\pi\beta ni}g_{n}(\Delta r),
\end{eqnarray}
where
\begin{eqnarray}\label{eq18WFQuasiperiodicas2}
g_n(\Delta r)= \frac{1}{\left[\left(\Delta x - an\right)^2 + \Delta y^2 +  \Delta z^2 -  \Delta t^2\right]}.
\end{eqnarray}
This is the positive frequency Wightman function for the massless scalar field subjected to a quasiperiodic condition. The Minkowski divergent contribution can now be easily separated to be subtracted in the renormalization process, leading to finite renormalized velocity dispersions. This term again arises from the $n=0$ contribution of the above sum.
Note that in the periodic case, $\beta=0$, Eq. \eqref{eq18WFQuasiperiodicas} corresponds to a spacetime of topology $S^{1}\times R^{3}$, that is, a compactified direction in a circle and a three-dimensional space of coordinates $(t,y,z)$ with $t\geq 0$, $-\infty<y<\infty$ and $-\infty<z<\infty$. This spacetime configuration is shown in Fig.\ref{figR3xS1}. Then, the case $\beta\neq 0$ can be thought of as a generalization, which we called modified $S^{1}\times R^{3}$ spacetime, because of the phase introduced by the quasiperiodic parameter $\beta$.

With the convenient form for the Wightman functions obtained in this section we can proceed to the next section to calculate the renormalized velocity dispersion in each condition scenario.

%
%%%%%%%%%%%%%%%%%%%%%%%%%%%%
\section{Velocity despersions}\label{SecVelDis}
%%%%%%%%%%%%%%%%%%%%%%%%%%%%
\subsection{General expression}
%%%%%%%%%%%%%%%%%%%%%%%%%%%%
%
Let us now analyze the dynamics of the point particle coupled to the massless scalar field. Thus, by varying the action \eqref{eq01} with respect to the position, we obtain the following expression for particle's velocity \cite{camargo2019vacuum,camargo2018vacuum,Camargo:2020fxp, de2014quantum}:
\begin{eqnarray}\label{eq19}
v_{i}(\tau,{\bf x})=-\frac{g}{m}\int_{0}^{\tau}dt\frac{\partial\phi({\bf x},t)}{\partial x_{i}},
\end{eqnarray}
where $i=(x,y,z)$ and we have considered a null initial velocity, that is, $v_{i}(t=0)=0$. In the above expression, a more accurate description should take into account the fact
that the spatial coordinates are functions of time, i.e., $\text{{\bf x}}\equiv \text{{\bf x}}(t)$. However, we assume here that the particle's displacement is small enough, so that possible time variations are negligible, providing that the spatial coordinates appearing in Eq. \eqref{eq19} are practically time independent \cite{yu2004vacuum,yu2004brownian,yu2006brownian,seriu2008switching,seriu2009smearing, de2016probing,de2019remarks,de2014quantum,camargo2018vacuum}. In Ref. \cite{de2014quantum}, for instance, this approximation has been discussed in the case of a point-like reflecting boundary and the authors have obtained the condition for which
the particle's displacement can be taken to be small. In Sec.\ref{Sec_Pos_disp} we shall examine this assumption and the necessary requirements for the validity of our results more closely.

As we are interested in studying the QBM induced by quantum vacuum fluctuations we should promote the scalar field to an operator as in Eq. \eqref{eq05qf} which, consequently, leads to the quantization of Eq. \eqref{eq19} as well. As a result, we note that $\langle 0|v_{i}|0\rangle\equiv \langle v_{i}\rangle=0$, that is, the velocity mean value of the particle due to the quantum vacuum fluctuations vanishes since, by definition, $a|0\rangle =0$ and $\langle 0|a^{\dagger}=0$.

Although the velocity mean value vanishes, the quantum vacuum fluctuations on the velocity can be calculated through the following expression for the renormalized velocity dispersion \cite{mota2020induced, ferreira2022quantum}:
\begin{eqnarray}\label{eq20}
\langle(\Delta v_{i})^{2}\rangle_{\text{ren}} = \lim_{x\rightarrow x'}\left[ \langle v_{i}(x)v_{i}(x')\rangle -\langle v_{i}(x)v_{i}(x')\rangle_{\text{div}} \right],
\end{eqnarray}
where we have introduced the notation $\langle 0|(\ldots)|0\rangle\equiv \langle(\ldots)\rangle$. Note that the Minkowski divergent contribution has been subtracted from the velocity dispersion, something that is standard in the renormalization process. 

From Eqs. \eqref{eq19} and \eqref{eq20} the renormalized velocity dispersion is formally given by
\begin{eqnarray}\label{eq21}
\langle(\Delta v_{i})^{2}\rangle_{\text{ren}} = \frac{g^2}{2m^2}\int_{0}^{\tau}dt'\int_{0}^{\tau}dt\frac{\partial^{2}G_{\text{ren}}^{(1)}(x,x')}{\partial x_{i}'\partial x_{i}},
\end{eqnarray}
where $G^{(1)}(x,x')=\langle\{\hat{\phi}(x),\hat{\phi}(x')\}\rangle$ is the Hadamard function that can be obtained from the positive frequency Wightman function by the relation $G^{(1)}(x,x')=2\text{Re}\,W(x,x')$  \cite{fulling1989aspects}. We should point out that the renormalized Hadamard function in Eq. \eqref{eq21} is obtained by subtracting the divergent Minkowski contribution present in the Wightman function already discussed in the previous section. We should also point out that in order to establish the above expression we have symmetrized the fields, a common procedure adopted in quantum field theory \cite{camargo2018vacuum,gour1999will}.

Next, we shall use the Wightman functions obtained in the previous section jointly with Eq. \eqref{eq21} to calculate the renormalized particle velocity dispersion corresponding to each boundary condition. 

%%%%%%%%%%%%%%%%%%%%%%%%%%%%%%%%%%%%%
\subsection{Dirichlet, Neumann and mixed boundary conditions}\label{veldispDNM}
%%%%%%%%%%%%%%%%%%%%%%%%%%%%%%%%%%%%%
%
Let us start by taking into consideration the velocity dispersion induced by Dirichlet, Neumann and mixed boundary conditions. To do this, we first consider the direction perpendicular to the planes, i.e., the $x$-direction. Thereby, from Eqs. \eqref{FWDNM} and \eqref{eq21}, after carrying out the integrals and derivatives operations, we find
\begin{eqnarray}\label{eq22}
\langle(\Delta v_{x})^{2}\rangle^{\text{(i)}}_{\text{ren}} &=& -\frac{g^{2}}{16\pi^{2}m^{2}a^{2}}\left[ 2\sum_{n=1}^{\infty} \gamma_{n}^{\text{(i)}} R(n,\tau_{a}) - \sum_{n=-\infty}^{\infty}\delta_{n}^{\text{(i)}}R(x_{a}-n,\tau_{a})\right],
\end{eqnarray}
where we have conveniently defined the dimensionless parameters $x_{a}=x/a$, $\tau_{a}=\tau/a$ and the function
\begin{eqnarray}\label{eq220a}
R(r,\tau_{a}) = P(r,\tau_{a})+Q(r,\tau_{a}),
\end{eqnarray}
with
\begin{eqnarray}\label{eq22a}
P(r,\tau_{a}) &=& \frac{2\tau_{a}^{2}}{r^{2}(4r^{2}-\tau_{a}^{2})},\nonumber\\
Q(r,\tau_{a}) &=& \frac{\tau_{a}}{2r^{3}}\ln\left(\frac{2r+\tau_{a}}{2r-\tau_{a}}\right)^{2}.
\end{eqnarray}
Note that in order to evaluate the integrals in Eq. \eqref{eq21} we have used the identity \cite{de2014quantum,camargo2018vacuum}
\begin{eqnarray}\label{eq22b}
\int_{0}^{\tau}dt'\int_{0}^{\tau}dt f(|t-t'|) = 2\int_{0}^{\tau}d\xi(\tau-\xi)f(\xi).
\end{eqnarray}
The plot for Eq. \eqref{eq22} is shown in Fig.\ref{fig01}, for distinct boundary conditions. In particular, the plot for mixed boundary condition of types DN and ND coincide when one takes the value $x_a=0.5$ and differ for other values. 

\begin{figure}
\includegraphics[scale=0.5]{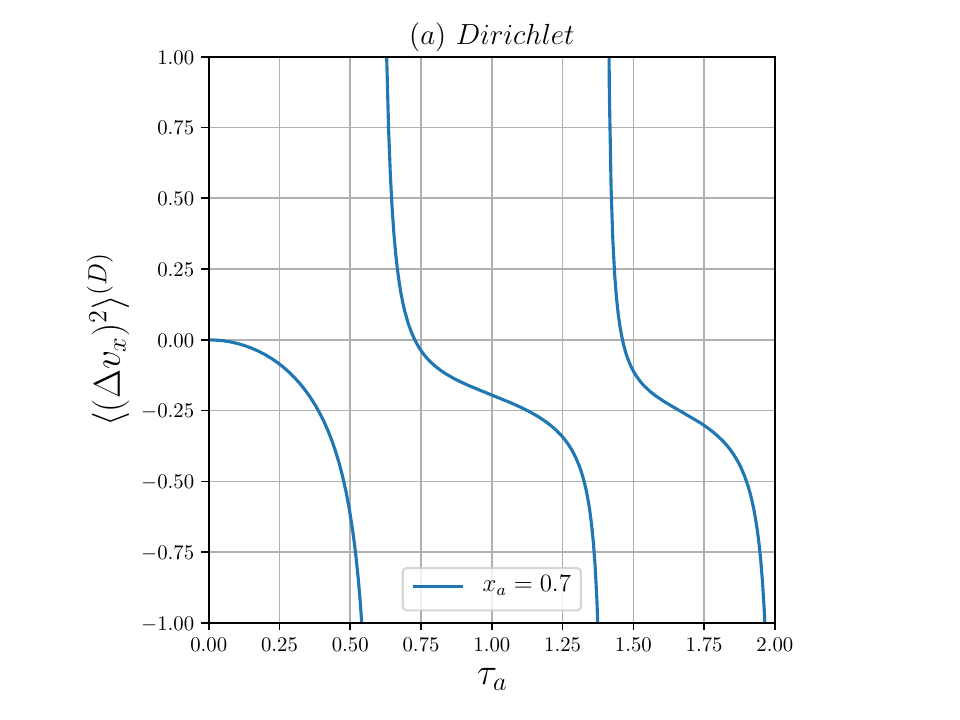}
\includegraphics[scale=0.5]{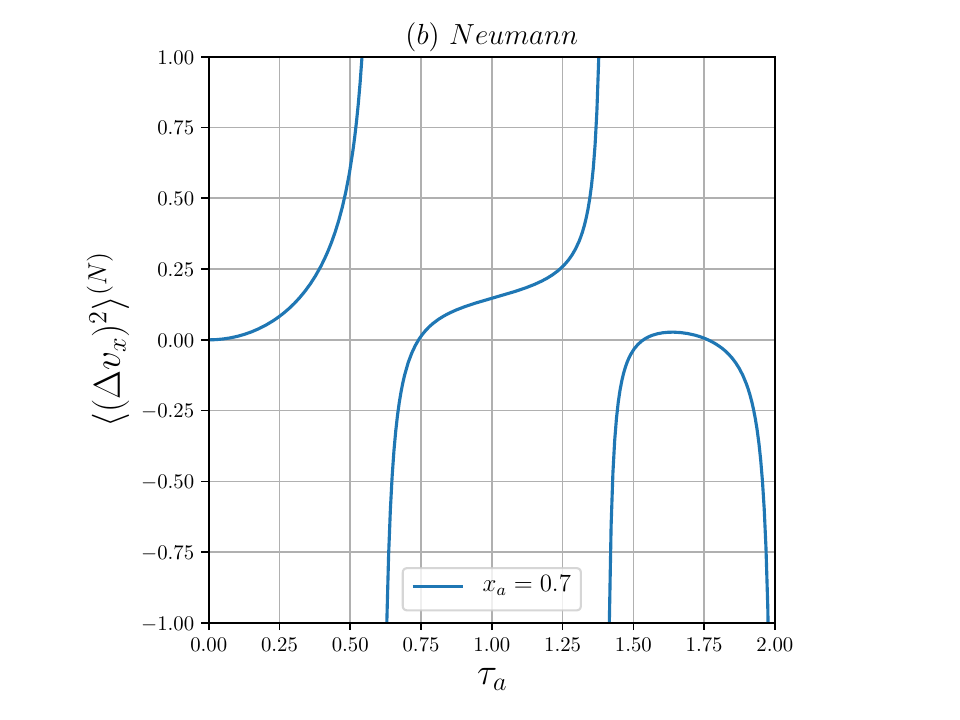}
\includegraphics[scale=0.5]{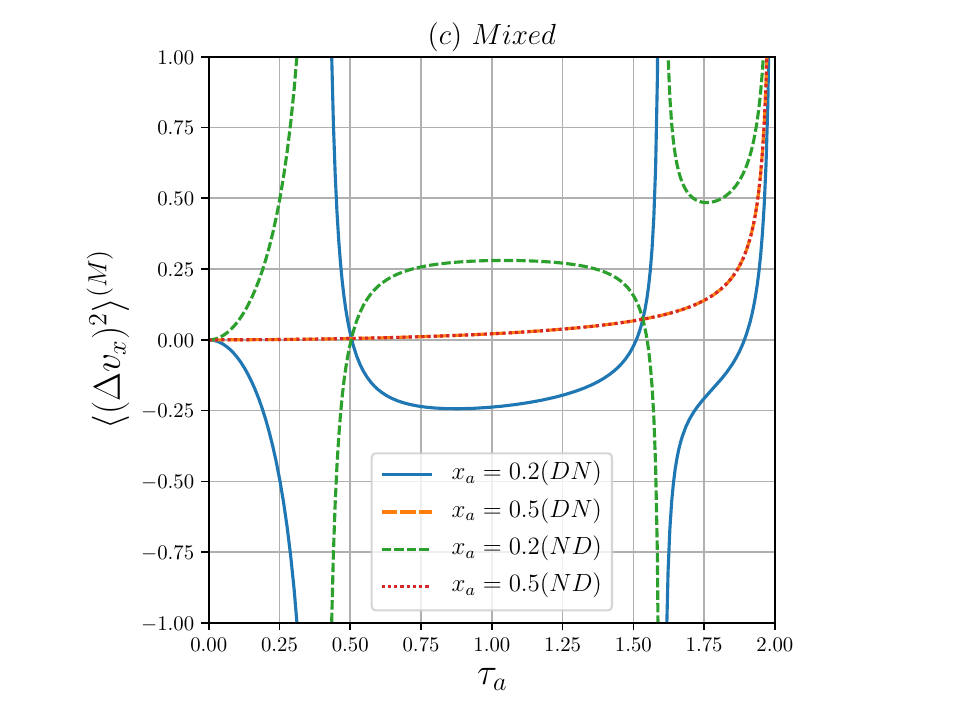}
\caption{Graph behavior of the perpendicular velocity dispersion for (a) Dirichlet (D), (b) Neumann (N) and (c) mixed (DN, ND) boundary conditions. Here we have considered the curves in units of $\langle(\Delta v_{x})^{2}\rangle^{\text{(i)}}=\langle(\Delta v_{x})^{2}\rangle^{\text{(i)}}_{\textrm{ren}}\left(\frac{ma}{g}\right)^{2}$. }\label{fig01}
\end{figure}
Similarly, for the velocity dispersion parallel to the planes we obtain 
\begin{eqnarray}\label{eq23}
\langle(\Delta v_{y})^{2}\rangle^{\text{(i)}}_{\text{ren}} &=& \dfrac{g^{2}}{32\pi^{2}m^{2}a^{2}}\left\{ 2\sum_{n=1}^{\infty}\gamma_{n}^{\text{(i)}}Q(n,\tau_{a}) + \sum_{n=-\infty}^{\infty}\delta_{n}^{\text{(i)}}Q(x_{a}-n,\tau_{a})  \right\},
\end{eqnarray}
where we have used the function $Q(r,\tau_a)$ defined in Eq. \eqref{eq22a}. The same result is obtained for the $z$ component of the velocity dispersion, also parallel to the planes. The behavior of this expression is depicted in Fig.\ref{fig02}. Again, in the case of mixed boundary condition, the plot shows that the curves for DN and ND coincide for $x_a=0.5$ and differ when taking other values.
\begin{figure}
\includegraphics[scale=0.5]{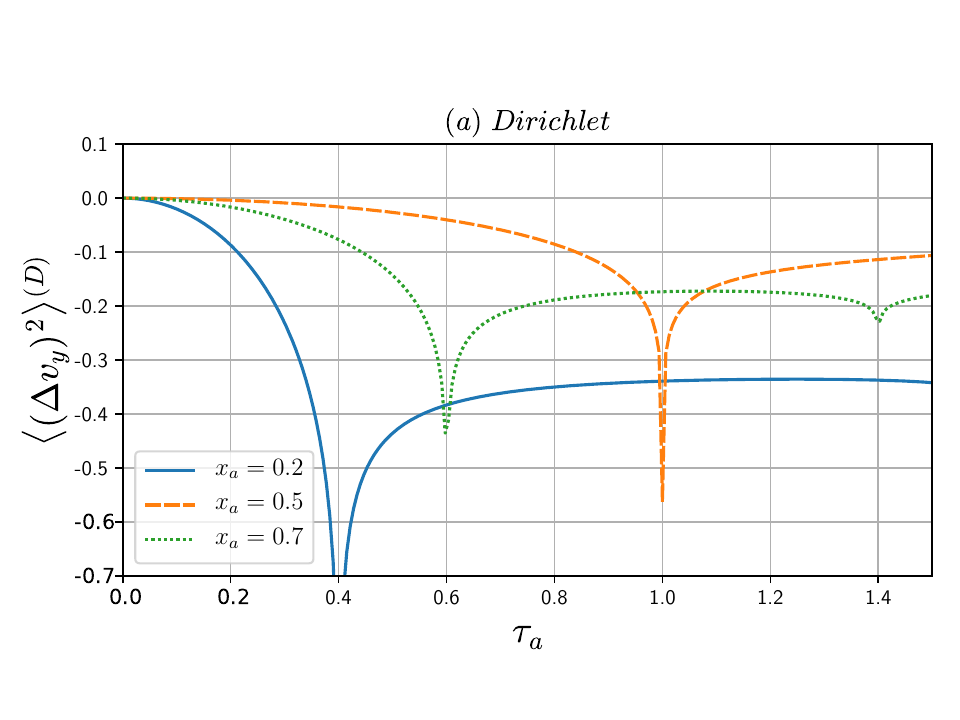}
\includegraphics[scale=0.5]{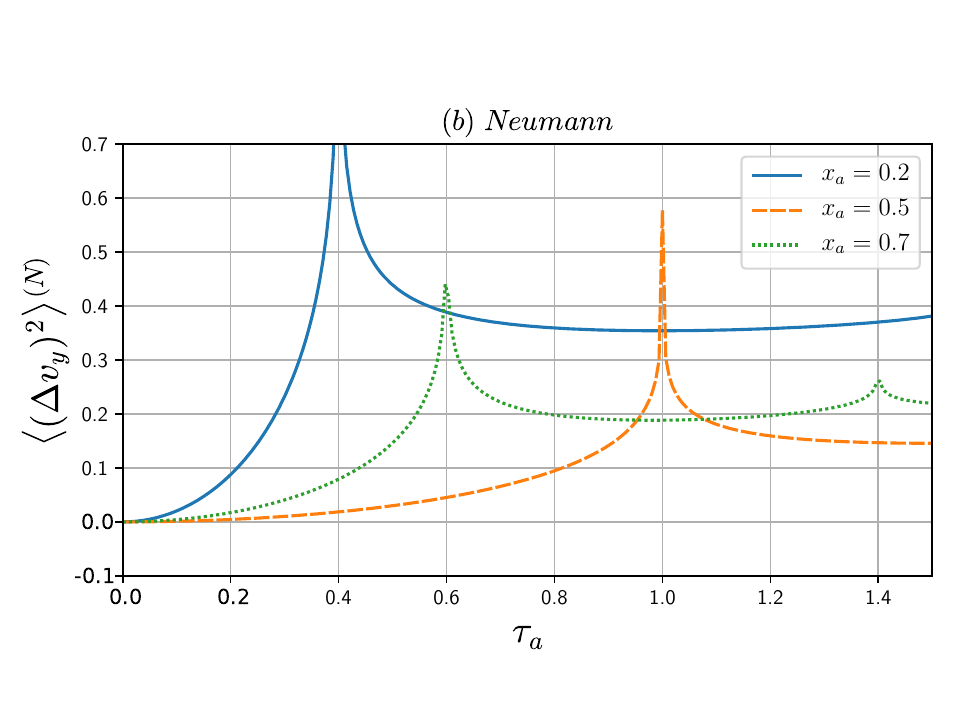}
\includegraphics[scale=0.5]{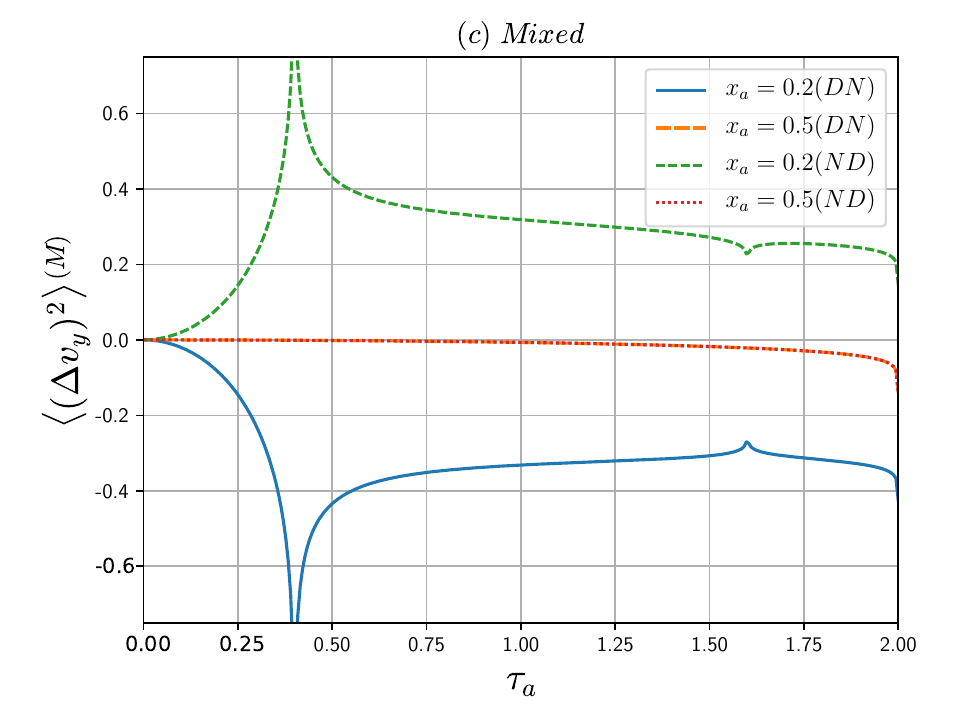}
\caption{Graph behavior of the parallel velocity dispersion curves for (a) Dirichlet (D), (b) Neumann (N) and (c) mixed (DN, ND) boundary conditions. Here we have considered the curves in units of $\langle(\Delta v_{y})^{2}\rangle^{\text{(i)}}=\langle(\Delta v_{y})^{2}\rangle^{\text{(i)}}_{\textrm{ren}}\left(\frac{ma}{g}\right)^{2}$. Note that the shown peaks represent divergent points.}\label{fig02}
\end{figure}

	It should be observed that, for Dirichlet boundary condition, the $n=0$ term of the expressions \eqref{eq22} and \eqref{eq23} corresponds to the one plane contribution for the velocity dispersions which has already been investigated in Ref. \cite{camargo2018vacuum}. This contribution is obtained from the second term in the r.h.s of Eq. \eqref{eq09WFDirichlet} for $n=0$. The latter, of course, is the Wightman function for Dirichlet boundary condition considering only one plane. Note that the one plane contribution for the velocity dispersions in the case of Neumann boundary condition is the same as the one for Dirichlet boundary condition, but with the opposite sign. Note also that the mixed boundary condition is not applicable for one single plane.

We now want to discuss the divergencies present in the expressions \eqref{eq22} and \eqref{eq23}. The first of them are the usual divergencies for points on the planes, at $x_{a}=0$ and  $x_{a}=1$. They come from the second term in the r.h.s of Eqs. \eqref{eq22} and \eqref{eq23} when $n=0$ and $n=1$, respectively. In addition, for $x_a\neq 0,1$, there also exist divergencies associated with the time a light signal takes to travel, in a round trip, from the planes to a point located at $x_{a}$ \cite{yu2004vacuum,yu2004brownian}. Mathematically this is given by $\tau_{a}=2|x_{a}-n|$, which tells us that each mode of the field contributes with a divergency.  Finally, there are also position independent divergencies in the form of $\tau=2na$ coming from the first term in the r.h.s of Eqs. \eqref{eq22} and \eqref{eq23}.  These divergencies represent an increasing number (with the field modes) of round trips from one plane to the other taken by a light signal. All these divergencies can be seen in the plots present in Figs.\ref{fig01} and \ref{fig02} for each boundary condition considered so far. For instance, in the plot for Dirichlet boundary condition shown in Fig.\ref{fig01}, the position independent divergency takes place for $\tau_a=2$ when $n=1$, while the position dependent divergencies take place for $\tau_a=1.4$ (when $n=0$) and $\tau_a=0.6$ (when $n=1$). Note that a larger range for $\tau_a$ would show additional divergencies. Note also that the same analysis can be reached for other values of $x_a$. In Ref. \cite{de2014quantum}, similar divergences have been studied in a one-dimensional model. The authors have shown that by assuming that the particle position fluctuates according to a Gaussian distribution the divergencies are smeared out. It has also been shown in Refs. \cite{de2016probing} and \cite{camargo2018vacuum} that implementation of switching functions can eliminate these typical divergences.

Let us now turn to the investigation of the behavior of the expressions \eqref{eq22} and \eqref{eq23} when $\tau_{a}\gg 1$ and $\tau_{a}\ll 1$, that is, for late and short time regimes, respectively. We start with the short time regime which indicates the behavior of the system in its initial moments of observation. In this sense, by considering the results of Appendix \ref{AppBDNMshorttime}, from Eqs. \eqref{eq22} and \eqref{eq23}, we obtain, for the perpendicular direction,
\begin{eqnarray}\label{eq24}
	\langle(\Delta v_{x})^{2}\rangle^{\text{(J)}}_{\text{ren}}\simeq - \dfrac{g^{2}\tau_{a}^{2}}{16\pi^{2}m^{2}a^{2}}\left\{ 3\zeta(4)-\delta^{(J)}\frac{\pi^{4}}{2}[2+\cos(2\pi x_{a})]\csc^{4}(\pi x_{a})\right\}
\end{eqnarray}
%_{\tau_{a}\ll 1} 
and	
\begin{eqnarray}\label{eq25}
	\langle(\Delta v_{x})^{2}\rangle^{\text{(M)}}_{\text{ren}} \simeq \dfrac{g^{2}\tau_{a}^{2}}{128\pi^{2}m^{2}a^{2}}\left\{ 21\zeta(4)-\delta^{(M)}\pi^{4}[11+\cos(2\pi x_{a})]\cot(\pi x_{a})\csc^{3}(\pi x_{a})\right\},
\end{eqnarray}	
while for the parallel direction we have
\begin{eqnarray}\label{eq26}
	\langle(\Delta v_{y})^{2}\rangle^{\text{(J)}}_{\text{ren}}\simeq \dfrac{g^{2}\tau_{a}^{2}}{32\pi^{2}m^{2}a^{2}}\left\{ 2\zeta(4)+\delta^{(J)}\frac{\pi^{4}}{3}[2+\cos(2\pi x_{a})]\csc^{4}(\pi x_{a})\right\}
\end{eqnarray}
and
\begin{eqnarray}\label{eq27}
	\langle(\Delta v_{y})^{2}\rangle^{\text{(M)}}_{\text{ren}} \simeq -\dfrac{g^{2}\tau_{a}^{2}}{128\pi^{2}m^{2}a^{2}}\left\{ 7\zeta(4)+\delta^{(M)}\frac{\pi^{4}}{3}[11+\cos(2\pi x_{a})]\cot(\pi x_{a})\csc^{3}(\pi x_{a})\right\},
\end{eqnarray}	
where $\delta^{\text{(J)}} = [\delta^{\text{(D)}},\delta^{\text{(N)}}]=[-1,+1]$ and $\delta^{\text{(M)}} = [\delta^{\text{(DN)}},\delta^{\text{(ND)}}]=[+1,-1]$. We can see that all the expressions above for the short time regime are of order $\tau_a^2$, the leading order of Eqs. \eqref{apBeq07}, \eqref{apBeq09}, \eqref{apBeq18} and \eqref{apBeq20} considered to perform the analysis. Note that in the expressions above only the divergencies on the planes, at $x_{a}=(0,1)$, are preserved. 

On the other hand, similarly to the classical Brownian motion for a point particle immersed in a fluid at finite temperature, the late time regime in our case, that is, $\tau_a\gg 1$, also gives us an approximately time independent value for the velocity dispersion. In fact, based on the results of Appendix \ref{AppADNMlatetime}, this is shown in the expression below for the perpendicular direction, that is,
\begin{eqnarray}\label{eq28}
\langle(\Delta v_{x})^{2}\rangle^{\text{(J)}}_{\text{ren}}  \simeq -\frac{g^{2}}{8\pi^{2}m^{2}a^{2}}\left[\frac{\pi^{2}}{3}+\frac{4}{3\tau_{a}^{2}}-\delta^{(J)}\pi^{2}\csc^{2}(\pi x_{a})\right]
\end{eqnarray}
%_{\tau_{a}\gg 1}
and
\begin{eqnarray}\label{eq29}
\langle(\Delta v_{x})^{2}\rangle^{\text{(M)}}_{\text{ren}}  \simeq \frac{g^{2}}{8\pi^{2}m^{2}a^{2}}\left[\frac{\pi^{2}}{6}-\frac{4}{3\tau_{a}^{2}}-\delta^{(M)}\pi^{2}\cot(\pi x_{a})\csc(\pi x_{a})\right],
\end{eqnarray}
while for the parallel direction we have
\begin{eqnarray}\label{eq30}
\langle(\Delta v_{y})^{2}\rangle^{\text{(J)}}_{\text{ren}}  \simeq \frac{g^{2}}{8\pi^{2}m^{2}a^{2}}\left[\frac{\pi^{2}}{3}-\frac{4}{3\tau_{a}^{2}}+\delta^{(J)}\pi^{2}\csc^{2}(\pi x_{a})\right]
\end{eqnarray}
and
\begin{eqnarray}\label{eq31}
\langle(\Delta v_{y})^{2}\rangle^{\text{(M)}}_{\text{ren}} \simeq -\frac{g^{2}}{8\pi^{2}m^{2}a^{2}}\left[\frac{\pi^{2}}{6}+\frac{4}{3\tau_{a}^{2}}+\delta^{(M)}\pi^{2}\cot(\pi x_{a})\csc(\pi x_{a})\right],
\end{eqnarray}
where the coefficients $\delta^{\text{(J)}}$ and $\delta^{\text{(M)}}$ have already been previously defined below Eq. \eqref{eq27}. From Eqs. \eqref{eq28}, \eqref{eq29}, \eqref{eq30} and \eqref{eq31}, we see that all the expressions have a term of order $4/3\tau_{a}^{2}$, which is negligible for large time values so that the remainder terms are the dominant ones. In particular, the position dependent term depends on the boundary condition used and also preserves the divergencies on the planes located at $x_{a}=(0,1)$. 

%The velocity dispersions in Eqs. \eqref{eq22} and \eqref{eq23} becoming time independent at much later times is analogous to what happens in the classical Brownian motion for a point particle immersed in a fluid at finite temperature, which also becomes time independent in this regime.
%

We can show that the expressions for the late time regime obtained above, for Dirichlet boundary condition, when $x_{a}\ll 1$, is consistent with the result presented in Ref. \cite{camargo2018vacuum} where the authors considered a single plane. Thereby, expanding Eqs. \eqref{eq28} and \eqref{eq30} for $x_{a}\ll 1$ we found
\begin{equation}\label{oneplaneasymptotics}
\langle(\Delta v_{x})^{2}\rangle^{\text{(D)}}_{\text{ren}} = \langle(\Delta v_{y})^{2}\rangle^{\text{(D)}}_{\text{ren}}\simeq-\frac{g^{2}}{8\pi^{2}m^{2}x^{2}},
\end{equation}
%_{\tau_{a}\gg 1}
which is exactly Eq. (4.3) of Ref. \cite{camargo2018vacuum}. The limit $x_{a}\ll 1$ is equivalent to say that the plane placed at $x=a$ is moved far away from the plane at $x=0$, ideally to infinity (see Fig.\ref{figtwoparallelplanesB}). Consequently, the infinitely distant plane has no effect on the particle. Thus, the resulting scenario is a point particle in the presence of a single plane, placed at $x=0$, which is one of the configurations studied in Ref. \cite{camargo2018vacuum} for the late time regime.

In the case of mixed boundary condition we can observe a similar situation in the limit $x_{a}\ll 1$, that is, $a\rightarrow\infty$. In this case, we can show that the expressions for the velocity dispersion, Eqs. \eqref{eq25}, \eqref{eq27}, \eqref{eq29} and \eqref{eq31}, correspond to either Dirichlet or Neumann boundary condition only, depending whether we consider the DN or ND configuration on the planes. For instance,  let us consider the DN configuration, where $\delta^{(M)}\equiv\delta^{(DN)}=+1$. This is the configuration in which Dirichlet and Neumann boundary condition are applied to the planes placed at $x=0$ and $x=a$, respectively. In this sense, taking the limit $a\rightarrow\infty$ in the aforementioned expressions we obtain the result for Dirichlet boundary condition in the corresponding limit, namely, Eqs. \eqref{eq24}, \eqref{eq26}, \eqref{eq28} and \eqref{eq30}, with $\delta^{(J)}\equiv\delta^{(D)}=-1$. The explanation is that once we move the plane placed at $x=a$ to infinity, only the plane with Dirichlet boundary condition, at $x=0$, produces some effect on the particle. The argument for the ND configuration is similar. 
\begin{figure}[h]
\includegraphics[scale=0.2]{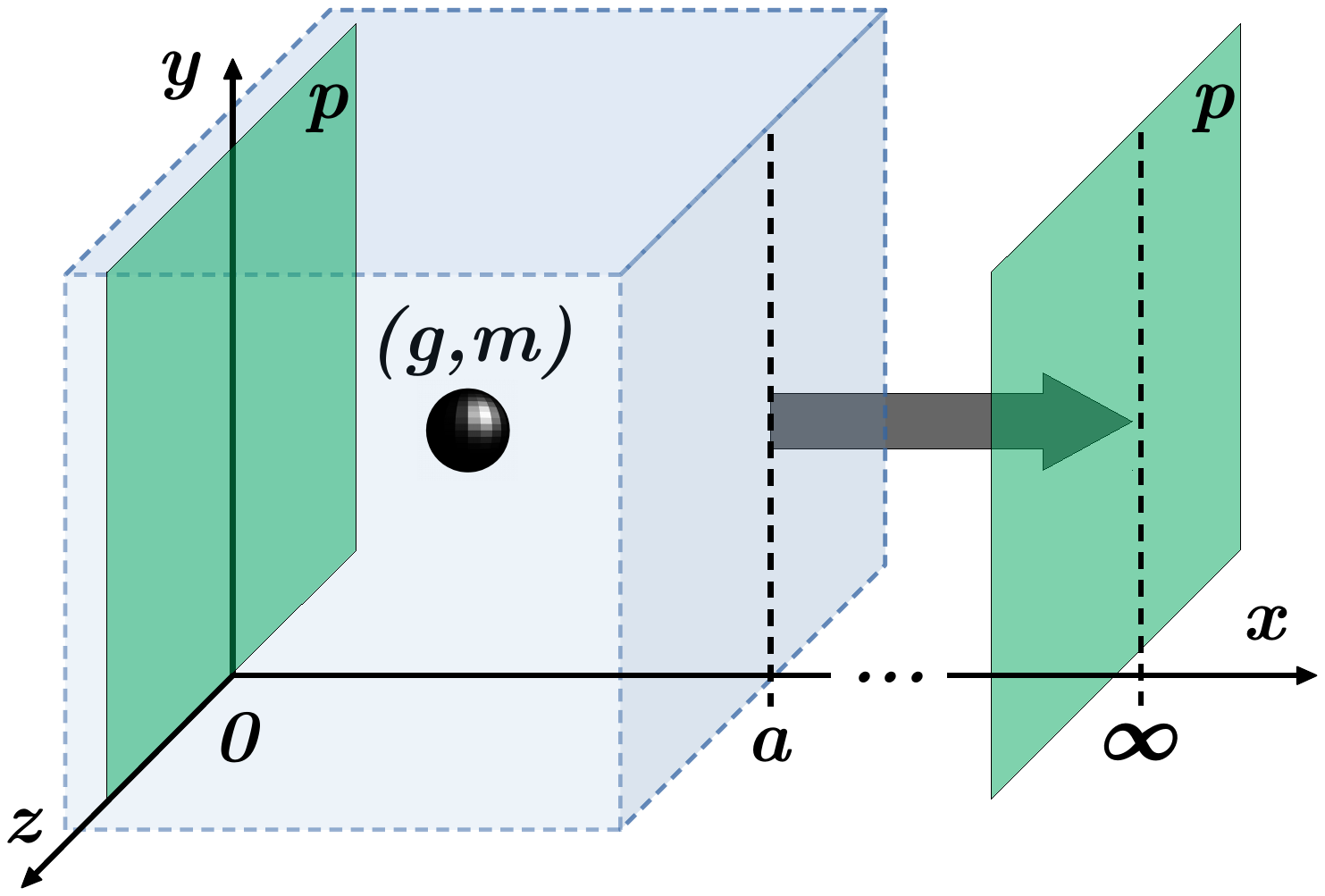}
\caption{If we move the plane ($p$) at $x=a$ to infinity everything works like if the plane at $x=a$ did no exist, that is, the resulting configuration is equivalent to a point particle in the presence of a single plane.}\label{figtwoparallelplanesB}
\end{figure}

Finally, to end this subsection we would like to make a brief comment about possible negative values that the velocity dispersions can take. This can be seen from Eq. \eqref{eq20}, which consists of a diference between the dispersion in the presence of two parallel planes and the dispersion without planes, which is divergent. So, a negative value indicates that the presence of the planes creates a reduction in the velocity dispersion, as argued in Ref. \cite{yu2004vacuum}. 
%
%

%%%%%%%%%%%%%%%%%%%%%%%%%%%%%%%%%%%%%%
\subsection{Quasiperiodic condition}
%%%%%%%%%%%%%%%%%%%%%%%%%%%%%%%%%%%%%%

In order to obtain velocity dispersions corresponding to the quasiperiodic condition in Eq. \eqref{eq17} we make use of Eqs. \eqref{eq18WFQuasiperiodicas} and \eqref{eq21}. So, for the velocity dispersion in the $x$-direction, that is, the compactified direction, we find 
\begin{eqnarray}\label{eq32}
\langle(\Delta v_{x})^{2}\rangle^{\beta}_{\text{ren}} = -\dfrac{g^{2}}{\pi^{2}m^{2}a^{2}}\sum_{n=1}^{\infty}U(n,\beta,\tau_{a}),
\end{eqnarray}
%\cos(2\pi\beta n)\left[ \dfrac{\tau_{a}^{2}}{n^{2}(n^{2}-\tau_{a}^{2})} + \dfrac{\tau_{a}}{2n^{3}}\ln\left(\dfrac{n+\tau_{a}}{n-\tau_{a}}\right)^{2}\right]
while for the $y$-direction (or $z$), the uncompactified direction, we have
\begin{eqnarray}\label{eq33}
\langle(\Delta v_{y})^{2}\rangle^{\beta}_{\text{ren}} = \dfrac{g^{2}}{2\pi^{2}m^{2}a^{2}}\sum_{n=1}^{\infty}T(n,\beta,\tau_{a}),
\end{eqnarray}
%\dfrac{\cos(2\pi\beta n)}{n^{3}}\ln\left(\dfrac{n+\tau_{a}}{n-\tau_{a}}\right)^{2}
%
where we have defined the function 
\begin{eqnarray}\label{eq34}
U(n,\beta,\tau_{a}) = S(n,\beta,\tau_{a}) + T(n,\beta,\tau_{a}),
\end{eqnarray}
with
\begin{eqnarray}\label{eq34a}
S(n,\beta,\tau_{a}) &=& \frac{\tau_{a}^{2}\cos(2\pi\beta n)}{n^{2}(n^{2}-\tau_{a}^{2})}
\end{eqnarray}
and
\begin{eqnarray}\label{eq34b}
T(n,\beta,\tau_{a}) &=& \frac{\tau_{a}\cos(2\pi\beta n)}{2n^{3}}\ln\left(\frac{n+\tau_{a}}{n-\tau_{a}}\right)^{2}.
\end{eqnarray}
Note that to perform the integrals that have lead to the above expressions we have used again the identity \eqref{eq22b}. Similar to the previous cases, the compactification parameter $a$ provides a natural scale to the system, so that we are able to define the dimensionless time parameter $\tau_{a}$. It is important to call attention to the fact that the quasiperiodic condition has the particular periodic and antiperiodic condition cases given by, respectively, $\beta=0$ and $\beta=1/2$. From Eqs. \eqref{eq32} and \eqref{eq33} we observe that the expressions depend exclusively on the quasiperiodic parameter $\beta$, dimensionless time $\tau_{a}$ and length $a$. The graph behavior for these expressions is shown in the Fig.\ref{figcm01}.
A similar result has been obtained in Ref. \cite{Bessa:2019aar} for a point particle in the presence of a quantized electromagnetic field in a spacetime with spatial section of nontrivial topology, known as $E_{16}$ or slab topology, which is essentially defined by Eqs. \eqref{eq18WFQuasiperiodicas} and \eqref{eq18WFQuasiperiodicas2} for the periodic case $(\beta=0)$. 

\begin{figure}
\includegraphics[scale=0.5]{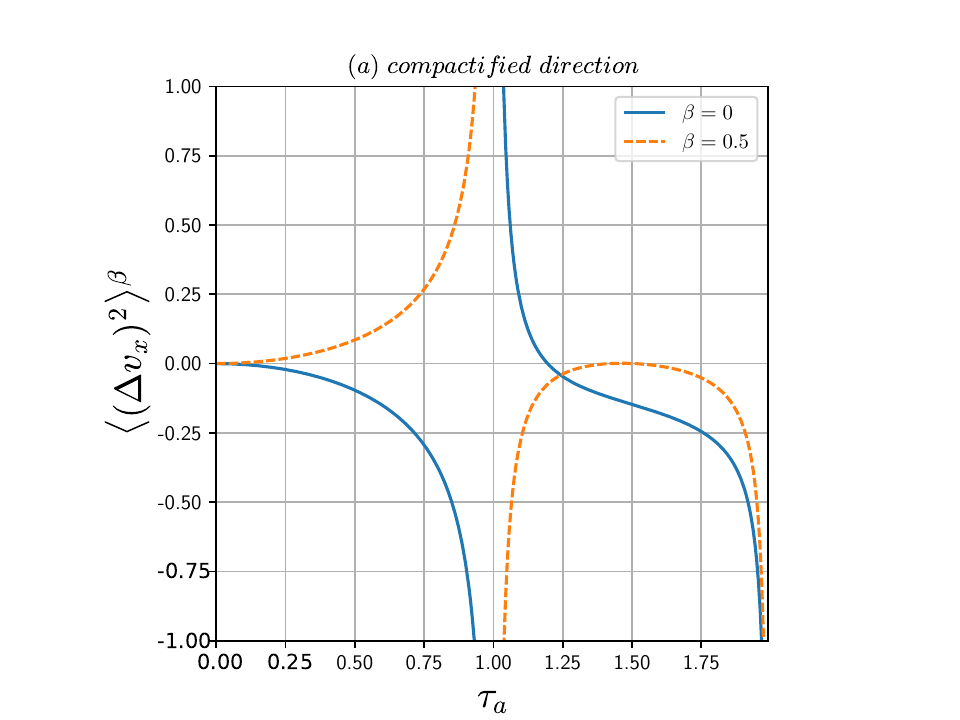}
\includegraphics[scale=0.5]{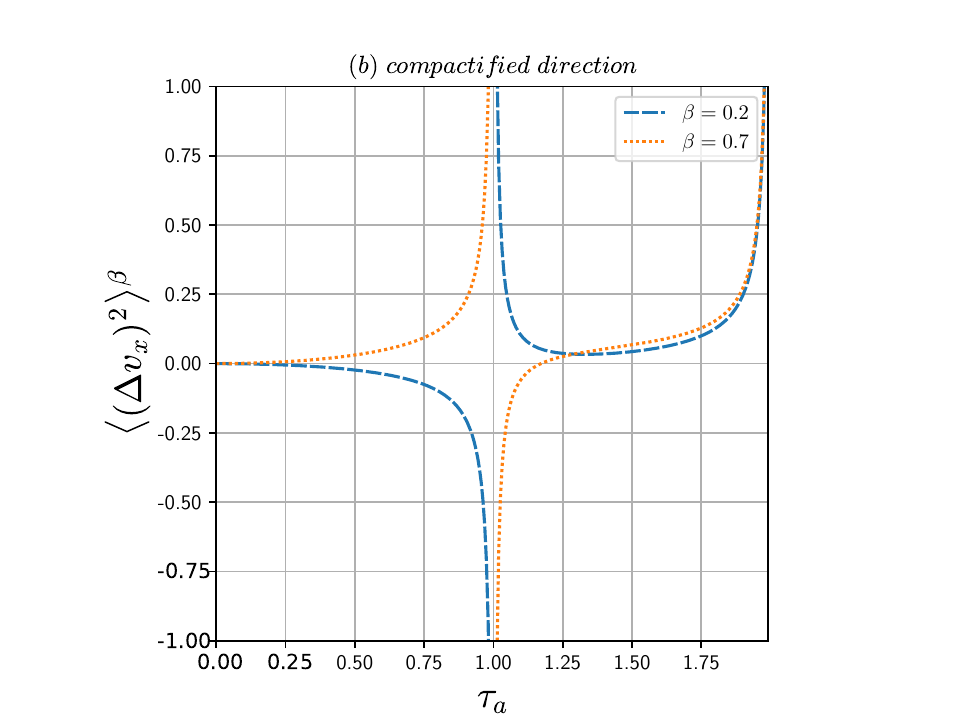}
\includegraphics[scale=0.5]{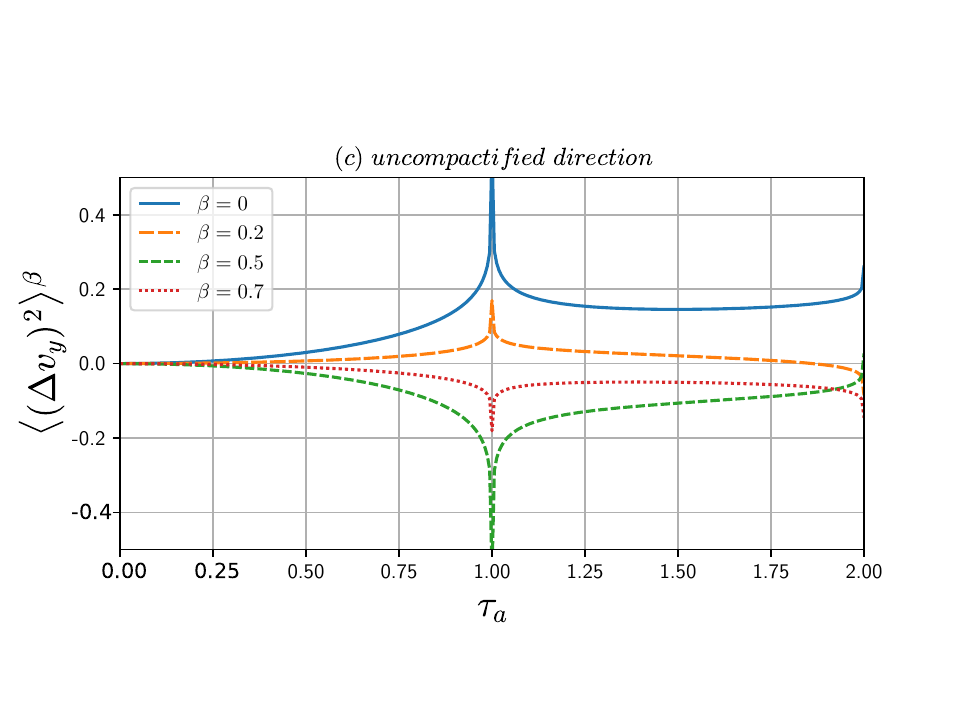}
\caption{Graph of the velocity dispersion in the (a)--(b) compactified and (c) uncompactified direction for the Quasiperiodic condition. Here we have considered the curves in units of $\langle(\Delta v_{x,y})^{2}\rangle^{\beta}=\langle(\Delta v_{x,y})^{2}\rangle^{\beta}_{\textrm{ren}}\left(\frac{ma}{g}\right)^{2}$. Note that the shown peaks represent divergent points.}\label{figcm01}
\end{figure}

	Differently from Dirichlet, Neumann and mixed boundary conditions, the expressions \eqref{eq32} and \eqref{eq33} do not have any dependency with the spatial coordinate $x$. The reason is that the quasiperiodic condition does not restrict the modes to a particular region as it happens to the parallel planes case. A spacetime in which one of the directions has a finite length $a$, as it is our case, makes possible to the modes to extend themselves throughout the whole $x$-coordinate. In contrast, in the Dirichlet, Neumann and mixed boundary condition cases the modes are also confined into a region of length $a$, but the $x$-component of the field does not exist outside this finite region. As we have already mentioned, the parallel planes break the homogeneity of the spatial section of the spacetime.
	
	Our expressions reveal that for some values of the time parameter $\tau_{a}$ we obtain divergent results to the velocity dispersion. Although the quasiperioric system is different, the interpretation of these singularities have similarity to those of the parallel planes case. Specifically, these divergencies occur for integer values of time, that is, $\tau_{a}=n$, as we can see from Eqs. \eqref{eq32} and \eqref{eq33}. The latter are plotted in Fig.\ref{figcm01} where, for the time range considered, there exist divergencies at $\tau_a=1$ and $\tau_a=2$. These divergencies are similar to the ones arising from the time a light signal takes to travel from a point $x_a$ to the planes in a round trip. However, in the quasiperiodic condition case, it is more intuitive to imagine circumferences of length $a$, so that $n$ values represent complete turns in the ciclic path. Then, we may understand the divergences in this case as due to the time taken by a light signal to travel an increasing number of ciclic paths of length $a$. As reported in Ref. \cite{Lemos:2020ogj}, where the authors analyzed the periodic case, the origin of such integer divergencies is a consequences of the spacetime topology, namely, $S^1\times R^3$.
%In addition, is instructive to check the results for the particulars cases periodic and antiperiodic

	Similarly to what has been done in the previous subsection, let us obtain the expressions for the short and late time regimes, that is, the velocity dispersions for the asymptotic time limits $\tau_{a}\ll 1$ and $\tau_{a}\gg 1$, respectively. From the results of Appendix \ref{AppBQPshorttime}, for the short time regime, the velocity dispersion in the $x$-direction is written as
\begin{eqnarray}\label{eq35}
\langle(\Delta v_{x})^{2}\rangle^{\beta}_{\text{ren}} \simeq \frac{g^{2}\tau_{a}^{2}\pi^{2}}{ m^{2}a^{2}}B_{4}(\beta), 
\end{eqnarray}
where $B_{n}(z)$ is the Bernoulli polynomial of order $n$ in the $z$ variable \cite{gradshtein2007}. The periodic (p) and antiperiodic (ap) cases are obtained as special cases of Eq. \eqref{eq35} for $\beta=0$ and $\beta=1/2$, respectively. These are given by
\begin{eqnarray}\label{eq35a}
\langle(\Delta v_{x})^{2}\rangle^{\text{(p)}}_{\text{ren}} \simeq -\frac{g^{2}\tau_{a}^{2}\pi^{2}}{30 m^{2}a^{2}}
\end{eqnarray}
and
\begin{eqnarray}\label{eq35b}
\langle(\Delta v_{x})^{2}\rangle^{\text{(ap)}}_{\text{ren}} \simeq\frac{7g^{2}\tau_{a}^{2}\pi^{2}}{240m^{2}a^{2}}.
\end{eqnarray}

Likewise, for the velocity dispersion in the $y$ (or $z$) direction, we find
\begin{eqnarray}\label{eq36}
\langle(\Delta v_{y})^{2}\rangle_{\text{ren}} \simeq -\dfrac{g^{2}\tau_{a}^{2}\pi^{2}}{3m^{2}a^{2}}B_{4}(\beta),
\end{eqnarray}
with
\begin{eqnarray}\label{eq36a}
\langle(\Delta v_{y})^{2}\rangle^{\text{(p)}}_{\text{ren}} \simeq \dfrac{g^{2}\tau_{a}^{2}\pi^{2}}{90m^{2}a^{2}}
\end{eqnarray}
and
\begin{eqnarray}\label{eq36b}
\langle(\Delta v_{y})^{2}\rangle^{\text{(ap)}}_{\text{ren}} \simeq-\dfrac{7g^{2}\tau_{a}^{2}\pi^{2}}{720m^{2}a^{2}},
\end{eqnarray}
for the periodic and antiperiodic cases. It is interesting to note that, similar to the cases studied in Section \ref{veldispDNM}, our expressions here also show a second order time dependency. From Eqs. \eqref{eq35} and \eqref{eq36}, we observe that the dispersion for the uncompactified direction is $-1/3$ of the result for the compactified one. 
Furthermore, the sign of the velocity dispersions in the short time regime is defined by the Bernoulli polynomials. In fact, as we can see in Fig.\ref{figBernoulli}, $B_{4}(\beta)$ assumes positive values in the range $r_{-}\leq\beta\leq r_{+}$, but it is negative for any other values of $\beta$, where $r_{\pm}=[1\pm (1-4n)^{1/2})]/2$, with $n=1/\sqrt{30}$, are the physical roots taking into consideration the condition $0\leq\beta<1$. 
For the periodic case ($\beta=0$), the compactified and uncompactified velocity dispersions achieve their minimum and maximum value, respectively, Eqs. \eqref{eq35a} and \eqref{eq36a}. On the other hand, in the antiperiodic case, Eqs. \eqref{eq35b} and \eqref{eq36b}, the opposite occurs.

\begin{figure}[h]
\includegraphics[scale=0.5]{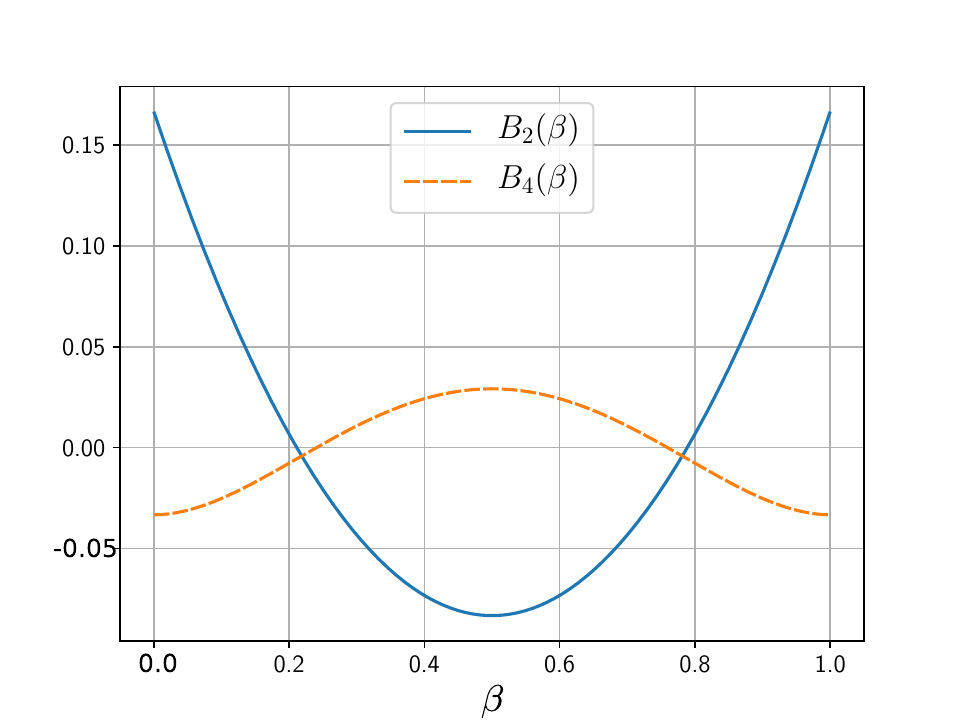}
\caption{Bernoulli polinomials $B_{2}(\beta)$, solid line, and $B_{4}(\beta)$, dashed line, as functions of the quasiperiodic parameter $\beta$.}\label{figBernoulli}
\end{figure}

We turn now to the analysis of the late time regime, that is, $\tau_{a}\gg 1$. Hence, by making use of the results in Appendix \ref{AppAQPlatetime}, for the $x$-direction, we have
\begin{eqnarray}\label{eq37}
\langle(\Delta v_{x})^{2}\rangle^{\beta}_{\text{ren}} \simeq -\dfrac{g^{2}}{\pi^{2}m^{2}a^{2}}\left[\pi^{2}B_{2}(\beta)+\frac{1}{6\tau_{a}^{2}}\right],
\end{eqnarray}
with
\begin{eqnarray}\label{eq37a}
\langle(\Delta v_{x})^{2}\rangle^{\text{(p)}}_{\text{ren}} \simeq-\dfrac{g^{2}}{6\pi^{2}m^{2}a^{2}}\left[\pi^{2}+\frac{1}{\tau_{a}^{2}}\right]
\end{eqnarray}
and
\begin{eqnarray}\label{eq37b}
\langle(\Delta v_{x})^{2}\rangle^{\text{(ap)}}_{\text{ren}} \simeq -\dfrac{g^{2}}{6\pi^{2}m^{2}a^{2}}\left[-\frac{\pi^{2}}{2}+\frac{1}{\tau_{a}^{2}}\right].
\end{eqnarray}
Additionally, for the $y$ (or $z$) direction, the velocity dispersion, in the late time regime, is given by
\begin{eqnarray}\label{eq38}
\langle(\Delta v_{y})^{2}\rangle^{\beta}_{\text{ren}} \simeq \dfrac{g^{2}}{\pi^{2}m^{2}a^{2}}\left[\pi^{2}B_{2}(\beta)-\frac{1}{6\tau_{a}^{2}}\right],
\end{eqnarray}
with
\begin{eqnarray}\label{eq38a}
\langle(\Delta v_{y})^{2}\rangle^{\text{(p)}}_{\text{ren}} \simeq\dfrac{g^{2}}{6\pi^{2}m^{2}a^{2}}\left[\pi^{2}-\frac{1}{\tau_{a}^{2}}\right]
\end{eqnarray}
and
\begin{eqnarray}\label{eq38b}
\langle(\Delta v_{y})^{2}\rangle^{\text{(ap)}}_{\text{ren}} \simeq-\dfrac{g^{2}}{6\pi^{2}m^{2}a^{2}}\left[\frac{\pi^{2}}{2}+\frac{1}{\tau_{a}^{2}}\right],
\end{eqnarray}
for the periodic and antiperiodic cases, respectively. The results above for the late time regime show that, from the last term in Eqs. \eqref{eq37} and \eqref{eq38}, the dispersions tend to a time independent value. Note that the contribution arising from the second term in the r.h.s of the expressions above for the late time regime is independent of the parameter $a$ and is identical for both compactified and uncompactified directions. Possibly, this suggests some kind of physical process which is independent of the compactification. In the compactified case, Eq. \eqref{eq37}, this small contribution tend to strengthen the dispersions whereas in the uncompactified case, Eq. \eqref{eq38}, it tends to weaken.

In the late time regime the quasiperiodic velocity dispersions can have a change of sign in the compactified and uncompactified cases. This is due to the behavior of $B_{2}(\beta)$ function shown in Fig.\ref{figBernoulli}. Moreover, only in the case where $\beta=(3\pm\sqrt{3})/6$ the time dependent small contribution define the sign of the velocity dispersions. Finally, we emphasize that the negative results for the velocity dispersions can be understood according to the interpretation given at the end of the previous subsection.
%
%
%

%%%%%%%%%%%%%%%%%%%%%%%%%%%%%%%%%%%%%%%%%%%%%%%%%%%%%%%%%%%%%%%%%%%%%%%%%
\subsection{Position dispersion and displacement condition}\label{Sec_Pos_disp}
%%%%%%%%%%%%%%%%%%%%%%%%%%%%%%%%%%%%%%%%%%%%%%%%%%%%%%%%%%%%%%%%%%%%%%%%%%
%
We want now to discuss a small displacement condition necessary to validate our results obtained for the velocity dispersions. Thus, we start by considering the expression
to calculate the dispersion of the position coordinates. Since $v=dx/dt$, the integration of Eq. \eqref{eq19} leads us to 
\begin{eqnarray}\label{disp_position_general_relation}
\langle(\Delta x_{i})^{2}\rangle^{\text{(j)}}_{\textrm{ren}} = \dfrac{g^{2}}{2m^{2}}\int_{0}^{\tau}dt\int_{0}^{\tau}dt'\int_{0}^{t}dt_{1}\int_{0}^{t'}dt_{2}\dfrac{\partial^{2}G^{(1)}_{\textrm{ren}}(x,x')}{\partial x'_{i}\partial x_{i}},
\end{eqnarray}
where we consider that the particle's position is initially zero, that is, $x_{i}(t=0)=0$. This is the general expression for the mean value in the vacuum state of the dispersion of the position coordinates $i = (x,y,z)$ of the particle, which is subject to distinct boundary conditions j=(N,D,DN,ND,$\beta$). In view of the relation $G^{(1)}(x,x')=2\text{Re}\, W(x,x')$, the Hadamard functions $G^{(1)}(x,x')$ can be obtained from the results of Sec.\ref{SecWF}. Next, we shall present the expressions for the position dispersions, obtained using Eq. \eqref{disp_position_general_relation}, in order to analyze the restrictions imposed on our results as a consequence of a small displacement condition.
%
%

%%%%%%%%%%%%%%%%%%%%%%%%%%%%%%%%%%%%%
\subsubsection{Dirichlet, Neumann and mixed boundary conditions}
%%%%%%%%%%%%%%%%%%%%%%%%%%%%%%%%%%%%%
%
Let us first consider the position dispersions for Dirichlet, Neumann and mixed boundary condition cases. Thus, from Eqs. \eqref{FWDNM}} and \eqref{disp_position_general_relation} we obtain that the dispersion referring to the perpendicular direction to the planes, that is, the $x$-direction, is given by
\begin{eqnarray}\label{position_disp_perp_direction}
\langle(\Delta x)^{2}\rangle^{\text{(j)}}_{\textrm{ren}}=\dfrac{g^{2}}{24\pi^{2}m^{2}}\left[2\sum_{n=1}^{\infty}\gamma_{n}^{\text{(j)}} A(n,\tau_{a})-\sum_{n=-\infty}^{\infty}\delta_{n}^{\text{(j)}} A(x_{a}-n,\tau_{a})\right],
\end{eqnarray}
while for the parallel direction, $y$ and $z$, is 
\begin{eqnarray}\label{position_disp_par_direction}
\langle(\Delta z)^{2}\rangle^{\text{(j)}}_{\textrm{ren}}=\langle(\Delta y)^{2}\rangle^{\text{(j)}}_{\textrm{ren}}=\dfrac{g^{2}}{24\pi^{2}m^{2}}\left[2\sum_{n=1}^{\infty}\gamma_{n}^{\text{(j)}} \bar{A}(n,\tau_{a})+\sum_{n=-\infty}^{\infty}\delta_{n}^{\text{(j)}} \bar{A}(x_{a}-n,\tau_{a})\right],
\end{eqnarray}
where for practical purposes we have defined the auxiliary functions
\begin{eqnarray}\label{position_disp_par_directionA}
A(r,\tau_{a}):= D(r,\tau_{a})+E(r,\tau_{a})+H(r,\tau_{a})
\end{eqnarray}
and 
\begin{eqnarray}\label{position_disp_par_directionB}
\bar{A}(r,\tau_{a}):= D(r,\tau_{a})+E(r,\tau_{a})-\dfrac{1}{2}H(r,\tau_{a})
\end{eqnarray}
with
\begin{eqnarray}\label{position_disp_par_directionD}
D(r,\tau_{a}) = \ln\left(\dfrac{\tau_{a}^{2}-4r^{2}}{4r^{2}}\right)^{2},
\end{eqnarray}
\begin{eqnarray}\label{position_disp_par_directionE}
E(r,\tau_{a}):=\dfrac{\tau_{a}^{2}}{2r^{2}},
\end{eqnarray}
and
\begin{eqnarray}\label{position_disp_par_directionH}
H(r,\tau_{a}):=-\dfrac{\tau_{a}^{3}}{4r^{3}}\ln\left(\dfrac{\tau_{a}+2r}{\tau_{a}-2r}\right)^{2}.
\end{eqnarray}
Similarly to the velocity dispersion expressions, in the above equations we use the dimensionless definition of time  $\tau_{a}=\tau/a$ and position $x_{a}=x/a$, in terms of the characteristic length $a$ of the system, which establishes the distance between the planes. The coefficients $\gamma^{(j)}_{n}$ and $\delta^{(j)}_{n}$ have already been defined in Eq. \eqref{Coefgd}.

From Eq. \eqref{position_disp_par_directionA} we note that the perpendicular position dispersion, Eq. \eqref{position_disp_perp_direction}, has a divergent behavior at $\tau_{a}=2n$, for the position independent term, and $\tau_{a}=2|x_{a}-n|$, for the position dependent term, which are the typical round trip divergences of the  corresponding velocity dispersions. On the other hand, according to Eq. \eqref{position_disp_par_directionB}, the parallel component of the position dispersion, Eq. \eqref{position_disp_par_direction}, is well defined at these same values. The finiteness of the position dispersion at the round trip values also happens in the case analyzed in Ref. \cite{de2014quantum} in (1+1) dimensions. However, in our case, as pointed out above, this only happens for the parallel components of the position dispersion. 

As we have mentioned in the paragraph bellow Eq. \eqref{eq19}, a precise and rigorous description of the velocity dispersion calculation must take into account the possible temporal dependence of the particle position coordinates. Nevertheless, we have assumed the hypothesis that such variations are negligible in order to obtain the analytical expressions for the velocity dispersion presented in Sec.\ref{veldispDNM}. This has an impact on the validation of our results and, as a consequence, an analysis to show a satisfactory margin of accuracy should be provided. 

The small displacements condition can be understood as the requirement that the modulus of the relative position dispersion be less than one. Mathematically, this is expressed as \cite{yu2004vacuum,de2014quantum}
\begin{eqnarray}\label{small_displacement_pp}
\dfrac{|\langle(\Delta x_{i})^{2}\rangle^{\text{(j)}}_{\textrm{ren}}|}{x^{2}}\ll 1.
\end{eqnarray}

From Eqs. \eqref{position_disp_perp_direction} and \eqref{position_disp_par_direction} we note that the free parameters in these expressions are the charge $g$, mass $m$, and the quantities $\tau_{a}$ and $x_{a}$. A straightforward way to check the small displacement condition requirements is to plot a graph of the dimensionless relative dispersion using Eqs. \eqref{position_disp_perp_direction} and \eqref{position_disp_par_direction} divided by $x^{2}$, as indicated in Eq. \eqref{small_displacement_pp}. For simplicity, we will consider the perpendicular component of the position dispersion in the discussion, but similar conclusions can be obtained using the parallel components.

\begin{figure}[h]
\includegraphics[scale=0.5]{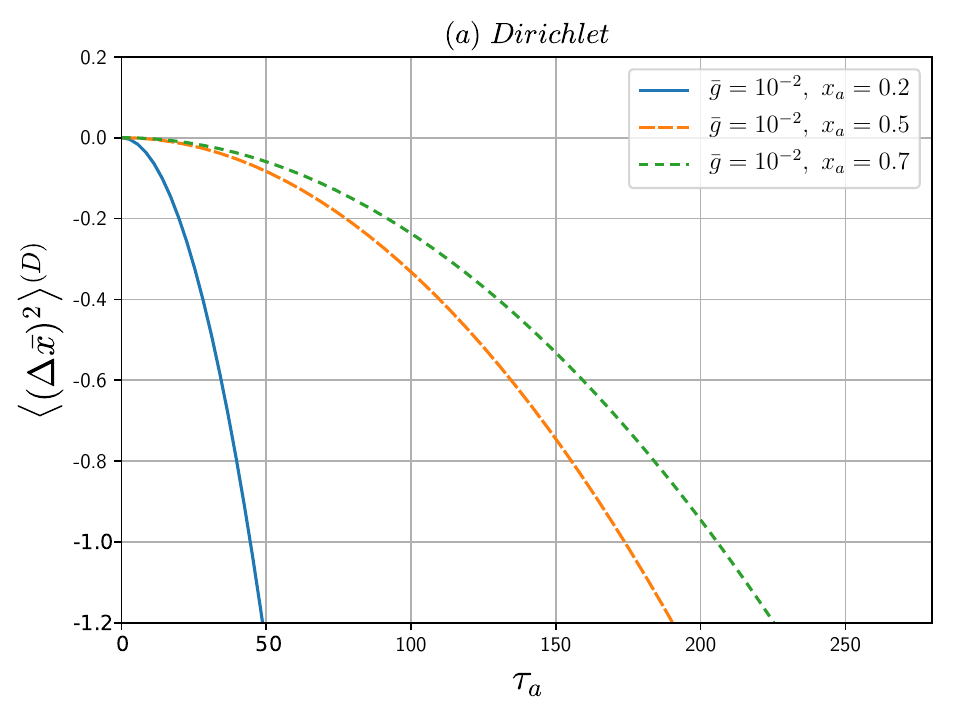}
\includegraphics[scale=0.5]{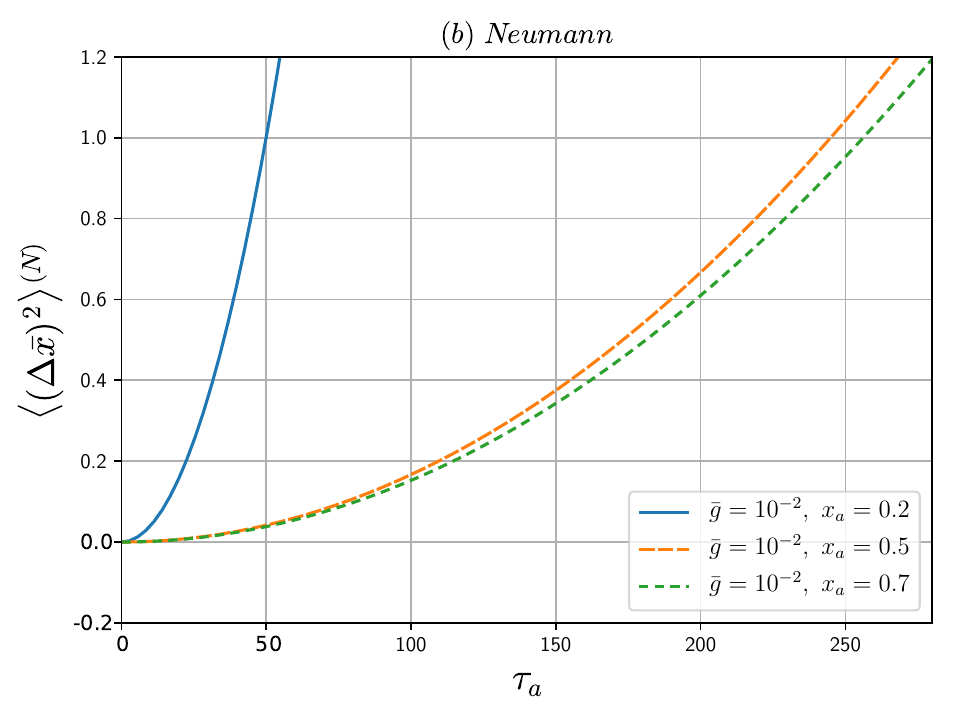}
\includegraphics[scale=0.5]{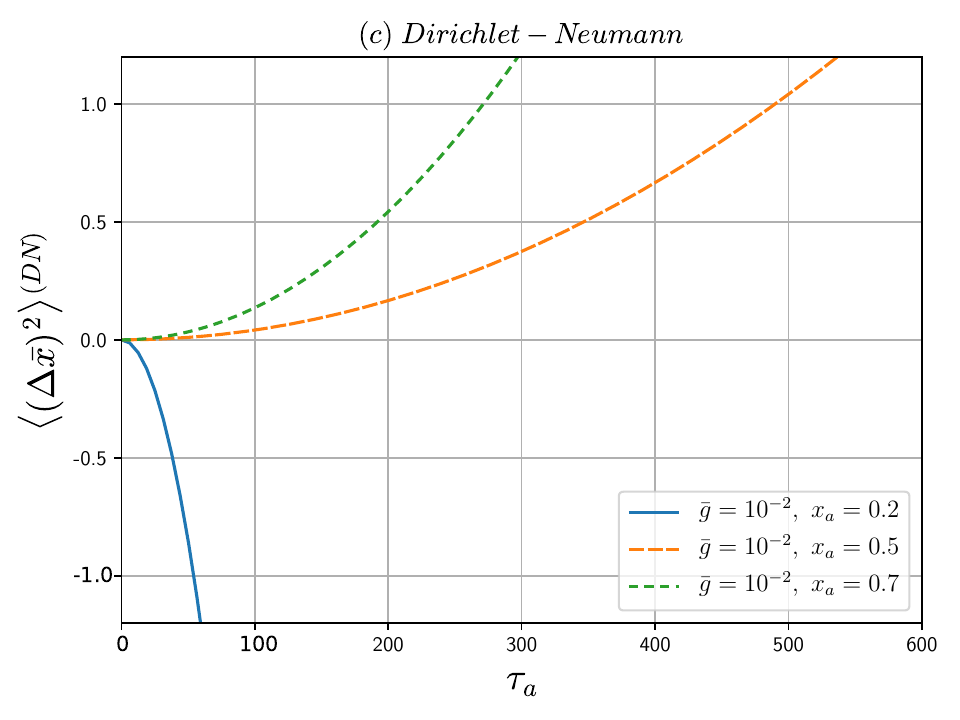}
\includegraphics[scale=0.5]{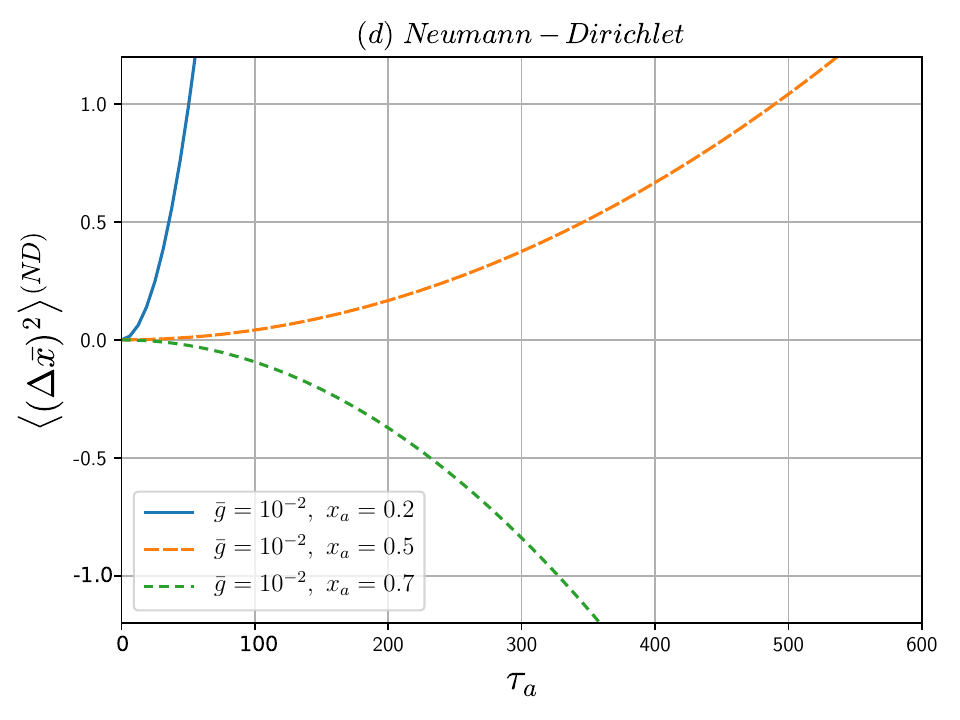}
\caption{Relative dispersion for the position coordinate perpendicular to the planes for (a) Dirichlet, (b) Neumann, (c) Dirichlet-Neumann and (d) Neumann-Dirichlet conditions, with $\bar{g}=\frac{g}{ma}=10^{-2}$ and different values of $x_{a}$. For simplicity, we have defined $\langle(\Delta \bar{x})^{2}\rangle^{\text{(j)}}:= \dfrac{\langle(\Delta x)^{2}\rangle^{\text{(j)}}_{\textrm{ren}}}{x^{2}} $.}\label{fig_disp_position_DNM}
\end{figure}

In Fig.\ref{fig_disp_position_DNM} we show the behavior of the relative position dispersion for the perpendicular direction, considering as an example $\bar{g}=\frac{g}{ma}=10^{-2}$ and different values for the relative position $x_{a}$. As we can see, the upper bound of condition \eqref{small_displacement_pp} is fixed by the intersection points between the relative dispersion curves and the straight horizontal lines $\pm 1$ in the figure grid. Therefore, suitable values of time $\tau_{a}$, in view of the small displacement condition, must occur before these intersection points. Taking this into consideration, the condition \eqref{small_displacement_pp} is satisfied with good effectiveness.

The upper bound values for the time $\tau_{a}$ verified graphically are shown in Part {\bf A} of Table \ref{tab1}. As we have pointed out, these values correspond to the intersection points between the relative position dispersion curves and the $\pm 1$ horizontal lines. Note that we have also considered $\bar{g}=10^{-3}$ in order to show the effect of $\bar{g}$ on the upper bound value for $\tau_{a}$. Based on this, we can notice that the smaller the value of $\bar{g}$, the greater the upper bound for $\tau_{a}$.

An analytical expression compatible with the upper bound values for $\tau_{a}$ shown in Table \ref{tab1} can be obtained from the late time regime of the perpendicular position dispersion, by making use of the results of Appendix \ref{AppADNMlatetime}. Thus, by following a similar methodology as the one used for the velocity dispersion, it is possible to obtain expressions for the position dispersion when $\tau_{a}\gg 1$. Hence, in the late time regime the condition \eqref{small_displacement_pp} leads to a condition in $\tau_a$, i.e.,
\begin{eqnarray}\label{small_displacement_pp1}
\tau_{a}^{\text{(j)}}\ll \dfrac{4\sqrt{3}x_{a}}{\bar{g}\sqrt{|h^{\text{(j)}}(x_{a})|}},
\end{eqnarray}
where for Dirichlet and Neumann conditions
\begin{eqnarray}
h^{\text{(J)}}(x_{a}) = -1+\delta^{\text{(J)}}3\csc^{2}(\pi x_{a})
\end{eqnarray}
and for mixed conditions
\begin{eqnarray}
h^{\text{(M)}}(x_{a}) = \dfrac{1}{2}-\delta^{\text{(M)}}3\cot(\pi x_{a})\csc(\pi x_{a}),
\end{eqnarray}
with $\delta^{\text{(J)}}=[\delta^{\text{(D)}},\delta^{\text{(N)}}]=[-1,+1]$ e $\delta^{\text{(M)}}=[\delta^{\text{(DN)}},\delta^{\text{(ND)}}]=[+1,-1]$. From Eq. \eqref{small_displacement_pp1} we immediately note that the upper bound value for $\tau_{a}$ is inversely proportional to the parameter $\bar{g}$, showing that the smaller the value of $\bar{g}$ the gretar the upper bound in $\tau_{a}$. The upper bound values obtained from the condition \eqref{small_displacement_pp1} are shown in Part {\bf B} of Tabela \ref{tab1}, from which we can infer a good agreement with the graph results exhibited in Part {\bf A}.
\begin{table}[h]
\caption{Upper bound values for the time $\tau_{a}^{\text{(j)}}$ in the case of the perpendicular position dispersion, subjected to the small displacements condition, considering different values for the dimensionless parameters $\bar{g}$ and $x_{a}$. Parts {\bf A} and {\bf B} of the table, respectively, show the results numerically obtained from the graphs in Fig.\ref{fig_disp_position_DNM} and the results using Eq. \eqref{small_displacement_pp1} for Dirichlet (D), Neuman (N), Dirichlet-Neumann (DN) and Neuman-Dirichlet (ND) boundary conditions.}\label{tab1}
\begin{tabular}{p{1cm}p{1cm}p{2cm}p{2cm}p{2cm}p{2cm}}
\toprule
&		  & \multicolumn{4}{c}{${\bf Part \ A}$} 		\\ \cmidrule{3-6}
$\bar{g}$ & $x_{a}$ & {\bf (D)} & {\bf (N)}	  & {\bf (DN)}	&  {\bf (ND)}\\ \midrule\midrule
$10^{-2}$ & $0.2$ & 44.52456	& 49.98398    & 54.24018	& 50.50610	  \\ 
$10^{-2}$ & $0.5$ & 173.81479	& 244.95202   & 490.01965	& 489.90280   \\ 
$10^{-2}$ & $0.7$ & 205.75056	& 256.18831   & 271.36713	& 327.39830	  \\
$10^{-3}$ & $0.2$ & 445.87440	& 499.89202   & 542.43711	& 505.11814	  \\
$10^{-3}$ & $0.5$ & 1,737.32398 & 2.449.49068 & 4,900.14736	& 4,898.98026 \\
$10^{-3}$ & $0.7$ & 2,056.86223	& 2.561.88441 & 2,713.66182	& 3,274.03633 \\ \cmidrule{3-6}
&		  &  \multicolumn{4}{c}{${\bf Part \ B}$} 		\\ \cmidrule{3-6}
$10^{-2}$ & $0.2$ & 44.52863	& 49.98932    & 54.24540    & 50.51258 		\\ 
$10^{-2}$ & $0.5$ & 173.20508	& 244.94897   & 489.89794	& 489.89794 	\\ 
$10^{-2}$ & $0.7$ & 205.23990 	& 256.18834   & 271.35622	& 327.40381 	\\ 
$10^{-3}$ & $0.2$ & 445.28630 	& 499.89321   & 542.45403	& 505.12585 	\\
$10^{-3}$ & $0.5$ & 1,732.05080 & 2,449.48974 & 4,898.97948	& 4,898.97948	\\
$10^{-3}$ & $0.7$ & 2,052.39909	& 2,561.88346 & 2,713.56226	& 3,274.03813 	\\
\bottomrule
\end{tabular} 
\end{table}

In Section \ref{veldispDNM}, we have analyzed the behavior of the velocity dispersions within the time range $\tau_{a}=[0,2]$ shown in the graphs in Figs.\ref{fig01} and \ref{fig02}. Let us consider as an example the configuration in which $\bar{g}=10^{-3}$ and $x_{a}=0.5$, i.e., the particle is located at the midpoint between the planes. In this case, the time interval $\Delta\tau_{a}=2$ corresponds approximately to $0.115\%$, $0.082\%$ and $0.041\%$ of the upper bound value for Dirichlet, Neumann and mixed (Dirichlet-Neumann and Neumann-Dirichlet) boundary conditions, respectively. In other words, the range $\Delta\tau_{a}=2$ for $\bar{g}=10^{-3}$ and $x_{a}=0.5$, considering distinct boundary conditions, is in a percentage range of approximately $0.04\%-0.12\%$. Hence, we clearly see that our results satisfy the small displacements condition \eqref{small_displacement_pp1} with a good margin of applicability. On the other hand, for $\bar{g}=10^{-2}$ and $x_{a}=0.5$, we can verify that the range $\Delta\tau_{a}=2$, for each of the boundary conditions, is in the percentage range $0.4\%-1.2\%$. Thus, we note that $\bar{g}$ plays a decisively role in the validity of the results. Finally, we conclude that a small value for $\bar{g}$ causes an increase in the confidence level of applicability of our results. In Ref \cite{de2014quantum} similar conclusions have been reported. 
%
%
%
%
%
%%%%%%%%%%%%%%%%%%%%%%%%%%%%%%%%%%%%%%%%%%%%%%%%%%%%%%%%%%%%%%%%%%%%%%%%
\subsubsection{Quasiperiodic condition case}
%%%%%%%%%%%%%%%%%%%%%%%%%%%%%%%%%%%%%%%%%%%%%%%%%%%%%%%%%%%%%%%%%%%%%%%%
%
Similarly to the previous case, from Eqs. \eqref{eq18WFQuasiperiodicas} and \eqref{disp_position_general_relation} we obtain that the position dispersion in the compactified $x$-direction is given by the expression 
\begin{eqnarray}\label{dispersion_compactified_direction}
\langle(\Delta x)^{2}\rangle^{(\beta)}_{\textrm{ren}} = \left(\dfrac{g}{\pi m}\right)^{2}\sum_{n=1}^{\infty}\cos(2\pi\beta n)\left[\dfrac{\tau_{a}^{2}}{6n^{2}} + \dfrac{1}{12}\ln\left(\dfrac{\tau_{a}^{2}-n^{2}}{n^{2}}\right)^{2} -\dfrac{\tau_{a}^{3}}{6n^{3}}\ln\left(\dfrac{\tau_{a}+n}{\tau_{a}-n}\right)^{2} \right],
\end{eqnarray}
whereas for the uncompactified coordinates $y$ and $z$ we have
\begin{eqnarray}\label{dispersion_uncompactified_direction}
\langle(\Delta z)^{2}\rangle^{(\beta)}_{\textrm{ren}} = \langle(\Delta y)^{2}\rangle^{(\beta)}_{\textrm{ren}} = \left(\dfrac{g}{\pi m}\right)^{2}\sum_{n=1}^{\infty}\cos(2\pi\beta n)\left[\dfrac{\tau_{a}^{2}}{6n^{2}} + \dfrac{1}{12}\ln\left(\dfrac{\tau_{a}^{2}-n^{2}}{n^{2}}\right)^{2} +\dfrac{\tau_{a}^{3}}{12n^{3}}\ln\left(\dfrac{\tau_{a}+n}{\tau_{a}-n}\right)^{2} \right].
\end{eqnarray}
We point out that the position dispersion in the compactified direction in Eq. \eqref{dispersion_compactified_direction} presents the typical round trip divergences at $\tau_{a}=n$, which also occur in the velocity dispersions. However, the position dispersions in the uncompactified directions in Eq. \eqref{dispersion_uncompactified_direction} are regular at $\tau_{a}=n$.

The compactification of a coordinate through condition \eqref{eq17} give us a natural measurement scale for the system, which is the compactification length $a$. Hence, the condition of small displacements assumed here can be expressed by means of the restriction 
\begin{eqnarray}\label{small_displacement_qpc0}
\left|\dfrac{\langle(\Delta x_{i})^{2}\rangle^{(\beta)}_{\textrm{ren}}}{a^{2}}\right| \ll 1,
\end{eqnarray}
where $i=x,y,z$. Note that, unlike Eq. \eqref{small_displacement_pp}, the relative position dispersion in Eq. \eqref{small_displacement_qpc0} is defined in terms of the parameter $a$ because the results referring to the quasiperiodicity condition are independent of any position coordinate. Thus, the only parameter with dimension of length available for comparison purposes is the compactification length $a$. 

\begin{figure}[h]
\includegraphics[scale=0.5]{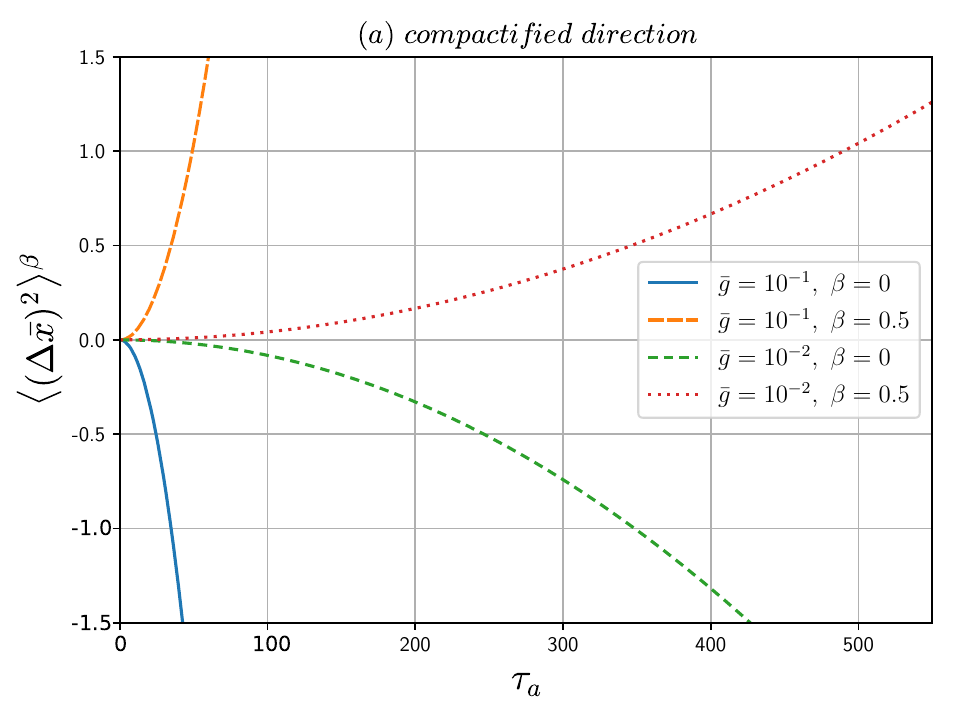}
\includegraphics[scale=0.5]{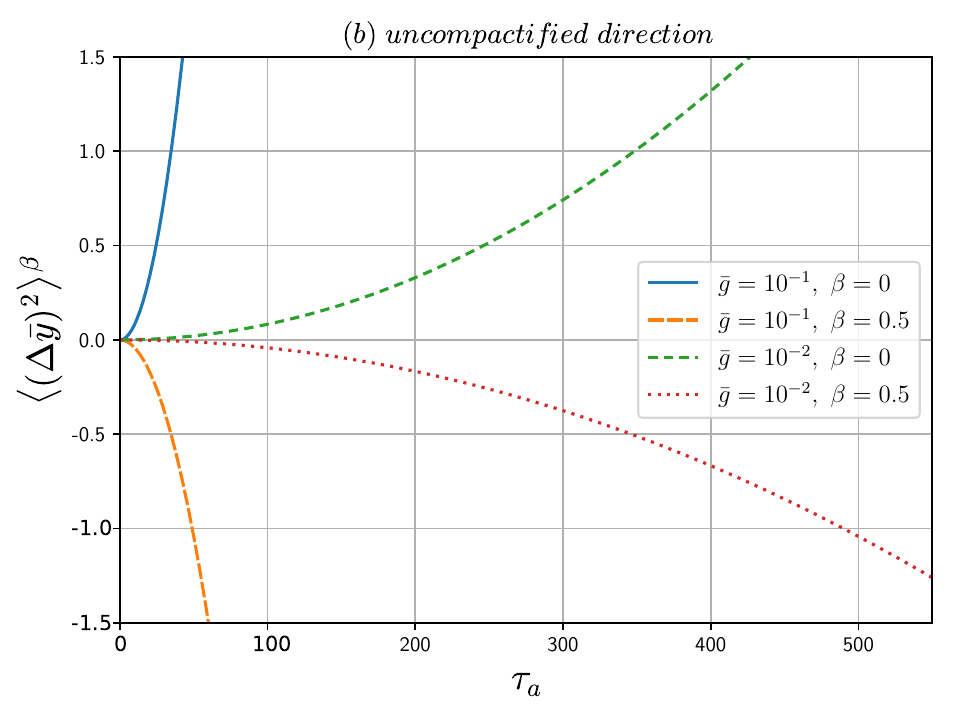}
\caption{Dimensionless position dispersion for (a) compactified and (b) uncompactified coordinates, with different values for the parameter $\bar{g}$ and the quasiperiodicity parameter $\beta$. Here we define $\langle(\Delta \bar{x}_{i})^{2}\rangle^{\beta}:= \dfrac{\langle(\Delta x_{i})^{2}\rangle^{\beta}_{\textrm{ren}}}{a^{2}} $.}\label{fig_disp_position1}
\end{figure}

The restrictions arising from the condition \eqref{small_displacement_qpc0} can be inferred  through a graph for the relative position dispersion using Eqs. \eqref{dispersion_compactified_direction} and \eqref{dispersion_uncompactified_direction}. Futhermore, as in the previous cases, to obtain a genuine numerical estimate of the magnitude of the relative position dispersion as a time function, it is necessary to define values for the dimensionless parameter $\bar{g}=g/ma$. Fig.\ref{fig_disp_position1} shows the dimensionless dispersions for the compactified and uncompactified position coordinates for two distinct values of $\bar{g}$, namely, $\bar{g}=10^{-1}$ and $\bar{g}=10^{-2}$. Observing the graphs it is possible to verify that the validity of Eq. \eqref{small_displacement_qpc0} is not violated for values of $\tau_{a}$ within the range defined by the intersection of the curves with the horizontal lines $\pm 1$. This, of course, also imposes an upper bound on the values of $\tau_{a}$.

In columns {\bf A} and {\bf B} of Table \ref{tab2}  we list the upper bound values for $\tau_{a}$ which satisfy condition \eqref{small_displacement_qpc0} for the relative dispersions in the compactified and uncompactified coordinate cases, respectively, for two values of parameters $\beta$ and $\bar{g}$. These values of $\tau_{a}$ correspond approximately to the points of intersection between the $\pm 1$ horizontal lines and the relative dispersion curves in Fig.\ref{fig_disp_position1}. 

It is possible to show that the the upper bound in Eq. \eqref{small_displacement_qpc0} can be be translated into an upper bound condition for the dimensionless time $\tau_a$ if one considers the late time regime. Hence, with the help of Appendix \ref{AppAQPlatetime} we obtain that the condition \eqref{small_displacement_qpc0} is satisfied for
\begin{eqnarray}\label{small_displacement_qpc1}
\tau_{a}\ll \dfrac{\sqrt{2}}{\bar{g}\sqrt{\left| B_{2}(\beta)\right|}}.
\end{eqnarray}
This expression clearly shows that the smaller the value of the parameter $\bar{g}$, the greater the upper bound on $\tau_a$, which consequently amplifies the range of applicability of our results. Column {\bf C} of Table \ref{tab2} displays the results obtained through Eq. \eqref{small_displacement_qpc1} for the upper bound values of the time $\tau_{a}$. Analyzing the results of columns {\bf A}, {\bf B} and {\bf C} of Table \ref{tab2} we note that Eq. \eqref{small_displacement_qpc1} provide us with an acceptable approximation for the upper bound values of $\tau_{a}$. 
%
%
%graphically observed
\begin{table}[h]
\caption{Values of the upper bounds for the dimensionless time $\tau_{a}$ numerically obtained from Fig.\ref{fig_disp_position1} for the compactified direction, column {\bf A}, uncompactified direction, column {\bf B}, and provided by condition \eqref{small_displacement_qpc1}, column {\bf C}.}\label{tab2}
\begin{tabular}{p{1cm}p{1cm}p{2cm}p{2cm}p{2cm}}
\toprule
		$\bar{g}$ & $\beta$ & {\bf A}   & {\bf B}	 & {\bf C}    \\ \midrule\midrule
		$10^{-1}$ & $0$	  	& 34.68574  & 34.67490   & 34.64101	\\ 
		$10^{-1}$ & $0.5$ 	& 49.01173  & 48.97372   & 48.98979	\\ 
		$10^{-2}$ & $0$	  	& 348.57922 & 348.41947  & 346.41016	\\ 
		$10^{-2}$ & $0.5$ 	& 489.95807 & 489.95444  & 489.89794	\\
\bottomrule
\end{tabular} 
\end{table}

In the graphs of the velocity dispersions the chosen range time is such that $\tau_{a}=[0,2]$, which can be seen in Fig.\ref{figcm01}. Considering the case $\bar{g}=10^{-2}$ as an example, it is possible to note that this range corresponds to less than $1\%$ of the upper bound imposed by condition \eqref{small_displacement_qpc1}, specifically, an approximate value of $0.6\%$ and $0.4\%$ in the periodic ($\beta=0$) and antiperiodic ($\beta=0.5$) cases for both directions, respectively. Therefore, the time range considered in the velocity dispersion graphs satisfactorily agrees with the small displacement assumption discussed above. Note again that the smaller the value of the parameter $\bar{g}$, the greater the margin of validity of our results.
%
%
%
%
%

%%%%%%%%%%%%%%%%%%%%%%%%%%%%%%%%
\section{Conclusions}\label{Conclusions}
%%%%%%%%%%%%%%%%%%%%%%%%%%%%%%%%
%
In this paper we have studied the QBM of a point particle induced by the quantum vacuum fluctuations of a massless scalar field, which are modified by both the presence of two reflecting parallel planes and a quasiperiodic condition that causes the $x$-direction to be compactified. We have considered three distinct boundary conditions for the field modes to obey on the planes placed perpendicular to the $x$-direction at $x=0$ and $x=a$. The boundary conditions are Dirichlet, Neumann, and  mixed which lead to the discretization of the momentum in the $x$-direction. Similarly, the quasiperiodic condition also leads to the discretization of the momentum in the $x$-direction in the form $k_{x} = k_{n} = \frac{2\pi}{a}(n+\beta)$, with $0\leq\beta < 1$. In all cases, the parameter $a$, related to confinement, provide a natural scale for the system, which enable us to analyze the resulting expressions in asymptotic regimes of interest, namely, short time ($\tau_{a}\ll 1$) and late time ($\tau_{a}\gg 1$) regimes. In the cases of Dirichlet, Neumann and mixed boundary conditions this parameter is the distance between the planes and for the quasiperiodic condition it is the quasiperiodicity length of the space or, in other words, the length of the compactification in the $x$-direction.

In the short time regime, for all conditions, we have seen that the most significant contributions for the velocity dispersions are of second order in time. For the late time regime, on the other hand, we have found that the velocity dispersions tend to a time independent expression, but which depends on the conditions imposed on the field. This fact is somewhat similar to what happens in the classical Brownian motion which, for sufficiently large observation of time, attains to a time independent expression given by $3k_{B}T/m$, where $k_{B}$ is Boltzmann constant, $T$ the temperature and $m$ the mass of the particle \cite{pathria}. This indicates a state of thermal equilibrium between the particle and the surrounding medium with temperature $T$. Hence, even in our simplified study, where temperature and dissipation effects are neglected, the results indicate a time independent expression to which the velocity dispersions of the particle tends. For this reason, we have drawn attention to a possible similarity between the induced QBM studied here, and the classical Brownian motion .
	
Divergent results for the velocity dispersions have also been identified, which are related to the usual divergencies on the planes at $x=0$ and $x=a$, to the time a light signal takes to travel in a round trip from one of the planes to a point $x$ and to the time a light signal takes to travel throughout the compactified direction. We have also indicated a position independent divergence in the parallel planes case that are related to an increasing number of round trips that a light signal takes to go from one plane to the other. Furthermore, negative velocity dispersions have also been shown to be possible and, based on discussions found in the literature, this can be understood as a reduction in the particle velocity dispersion due to the presence of the planes and the compactification mechanism.

	We would like to stress that two planes configuration has been considered in order to complement the investigations for the electromagnetic and scalar fields found in the literature.  Hence, the more remarkable contribution of this work has been the analysis of the QBM induced by the massless scalar field with distinct boundary conditions on the two parallel planes, which until now had not been done, besides Dirichlet conditions adopted only for one plane. In fact, all the works so far had focused on Dirichlet boundary condition. Also, the Dirichlet boundary condition has only been considered on two parallel planes in a system considering the electromagnetic field \cite{yu2004brownian}.

	The compact form for the positive frequency Wightman function in cartesian coordinates presented in Eq. \eqref{FWDNM} is very interesting because its structure makes possible to write the result for three boundary conditions into a single expression, namely, Dirichlet, Neumann and mixed boundary conditions. This structure is very useful since it allows to extract the divergent Minkowski contribution and, consequently, obtain other finite physical observables besides the velocity dispersion considered here. For instance, the mean value of field squared, $\langle\phi^{2}\rangle=\lim_{x'\rightarrow x}\langle\phi(x)\phi(x')\rangle$, and the mean value of the force squared that acts on the particle, $\langle F^{2}\rangle=\lim_{x'\rightarrow x}\langle F(x)F(x')\rangle$. In fact, as it can be easily checked, all these mentioned quantities depend on the Wightman function. 
	
We have assumed in our investigation the hypothesis that the charged particle does not significantly displace in time. As a consequence, the margin of applicability of our results is restricted to the small displacement condition given by Eqs. \eqref{small_displacement_pp} and \eqref{small_displacement_qpc0}. We, then, have shown in Sec.\ref{Sec_Pos_disp} that the results obtained for the velocity dispersions, along with the plots shown within the time range $\tau_a=[0,2]$, are in agreement with the small displacement condition. 

{\acknowledgments}

E.J.B.F would like to thank the Brazilian agency Coordination for the Improvement of Higher Education Personnel (CAPES) for financial support. E.M.B.G thanks financial support from the Brazilian agency National Council for Scientific and Technological Development (CNPq). H.F.S.M is partially supported by CNPq under grant N$\textsuperscript{o}$  311031/2020-0.
\\

\appendix 

\section{Late time regime}\label{AppA}

\subsection{Dirichlet, Neumann and mixed boundary conditions}\label{AppADNMlatetime}

In this first part of the appendix we go to investigate the expression $R(r,\tau_{a})$, Eq. \eqref{eq220a}, on late time regime, that is, $\tau_{a}\gg 1$. For the sake of clarity and in view of the fact that the parallel dispersion, Eq. \eqref{eq23}, is written only in terms of the $Q(r,t)$ function, Eq. \eqref{eq22a}, we shall develop each contribution from $R(r,\tau_{a})$ separately. Before proceeding it is useful and practical to define the following quantities:
\begin{eqnarray}\label{apAeq00}
R^{\text{(i)}}:= P^{\text{(i)}}+Q^{\text{(i)}}, 
\end{eqnarray}
and
\begin{eqnarray}\label{apAeq01}
R_{x_{a}}^{\text{(i)}}:= P_{x_{a}}^{\text{(i)}}+Q_{x_{a}}^{\text{(i)}},
\end{eqnarray}
with
\begin{eqnarray}\label{apAeq00a}
P^{\text{(i)}}:=\sum_{n=1}^{\infty}\gamma_{n}^{\text{(i)}}P(n,\tau_{a}),
\end{eqnarray}	
\begin{eqnarray}\label{apAeq00b}
Q^{\text{(i)}}:=\sum_{n=1}^{\infty}\gamma_{n}^{\text{(i)}}Q(n,\tau_{a}),
\end{eqnarray}
\begin{eqnarray}\label{apAeq01a}
P_{x_{a}}^{\text{(i)}}:=\sum_{n=-\infty}^{\infty}\delta_{n}^{\text{(i)}}P(x_{a}-n,\tau_{a}),
\end{eqnarray}	
\begin{eqnarray}\label{apAeq01b}
Q_{x_{a}}^{\text{(i)}}:=\sum_{n=-\infty}^{\infty}\delta_{n}^{\text{(i)}}Q(x_{a}-n,\tau_{a}),
\end{eqnarray}
where the index `i' indicates the boundary conditions, namely, i=(D, N, DN, ND). The functions as defined above separates the position independent contributions from the position dependent ones. Also, the functions $P(r,\tau_{a})$ and $Q(r,\tau_{a})$ are defined in Eq. \eqref{eq22a}, with the coefficients $\gamma_{n}^{(i)}$ and $\delta_{n}^{(i)}$ defined in Eq. \eqref{Coefgd}. In order to ensure the organization and make clearer the method used in our calculations, let us dedicate one subsection for the quantity $R^{\text{i}}$, which is position independent, and other for $R^{\text{(i)}}_{x_{a}}$, which is particle position dependent. In addition, to avoid overloading the descriptive text with excessive repetition of references, we emphasize that all relations used in manipulations of the expressions can be found in Refs. \cite{prudnikov1986integralsvol1, gradshtein2007}.

%%%%%%%%%%%%%%%%%%%%%%%%%%%%%%%%%%%
\subsubsection{Position independent term}
%%%%%%%%%%%%%%%%%%%%%%%%%%%%%%%%%%%
For the late time regime, i.e., $\tau_{a}\gg 1$, Eq. \eqref{apAeq00a} can be appropriately written in the form
\begin{eqnarray}\label{apAeq02}
P^{\text{(i)}}=-2\sum_{k=0}^{\infty}\left(\frac{2}{\tau_{a}}\right)^{2k}\sum_{n=1}^{\infty}\frac{\gamma_{n}^{\text{(i)}}}{n^{2-2k}},
\end{eqnarray}
where we have used a series expansion for the denominator of $P^{\text{(i)}}$. 

In the case of Dirichlet and Neumann boundary conditions $\gamma_{n}^{\text{(D)}} = \gamma_{n}^{\text{(N)}}=1$. Then, since these coefficients are independent of the summation index, we obtain that
\begin{eqnarray}\label{apAeq02DN}
P^{\text{(J)}} = -\dfrac{\pi^2}{3}+\dfrac{4}{\tau_{a}^2},
\end{eqnarray}
for the dominant terms, with J=(D,N). To establish the above result we have also used the relation%,$0.233(1)$ de \citep[p. 8]{gradshtein2007},
\begin{eqnarray}\label{apAprop00}
\sum_{k=1}^{\infty}\frac{1}{k^{p}}=\zeta(p),\qquad\qquad \text{Re}(p)>1,
\end{eqnarray}
and the fact that $\zeta(-2m) = 0$, where $m$ is a natural number.%, $9.542(3)$ \cite[1038]{gradshtein2007}.

For the mixed boundary conditions, observing that $\gamma_{n}^{\text{(DN)}} = \gamma_{n}^{\text{(ND)}}=(-1)^{n}$, from the Eq. \eqref{apAeq02}, we obtain
\begin{eqnarray}\label{apAeq02M}
P^{\text{(M)}} = \dfrac{\pi^2}{6}+\dfrac{4}{\tau_{a}^2},
\end{eqnarray}
for the dominant terms, with M=(DN,ND). To achieve the previous result we have used the relation%, $0.233(2)$ de \cite[p. 8]{gradshtein2007},
\begin{eqnarray}\label{apAprop01}
\sum_{k=1}^{\infty}\frac{(-1)^{k+1}}{k^{p}}=\left(1-2^{1-p}\right)\zeta(p),\qquad\qquad \text{Re}(p)>0,
\end{eqnarray}
in addition to the fact that $\zeta(-2m) = 0$, where $m$ is a natural number.

A similar procedure can be applied to the $Q^{\text{(i)}}$ function. First, we rewrite Eq. \eqref{apAeq00b} in the form
\begin{eqnarray}\label{apAeq03}
Q^{\text{(i)}}=\tau_{a}^{2}\sum_{k=1}^{\infty}\frac{1}{(2k-1)}\left(\frac{2}{\tau_{a}}\right)^{2k}\sum_{n=1}^{\infty}\dfrac{\gamma_{n}^{(i)}}{n^{4-2k}},
\end{eqnarray}
where we have used the series expansion%, $1.513(1)$ de \cite[p. 53]{gradshtein2007},
\begin{eqnarray}\label{apAprop02}
\ln\left(\frac{1+x}{1-x}\right) = 2\sum_{k=1}^{\infty}\frac{x^{2k-1}}{(2k-1)}, \qquad\qquad x^{2}<1.
\end{eqnarray}
Now, by using the same relations and properties introduced previously in the computations of $P^{\text{(i)}}$,  namely Eqs. \eqref{apAprop00} and \eqref{apAprop01}, we can easily obtain for the Dirichlet and Neumann boundary conditions
\begin{eqnarray}\label{apAeq03DN}
Q^{\text{(J)}} = \frac{2\pi^{2}}{3}-\frac{8}{3\tau_{a}^{2}}
\end{eqnarray}
and 
\begin{eqnarray}\label{apAeq03M}
Q^{\text{(M)}} = -\frac{\pi^{2}}{3}-\frac{8}{3\tau_{a}^{2}}
\end{eqnarray}
for the mixed boundary conditions.

From Eq. \eqref{apAeq00} and the results \eqref{apAeq02DN}, \eqref{apAeq02M}, \eqref{apAeq03DN}, \eqref{apAeq03M}, we can establish that
\begin{eqnarray}\label{apAeqRDN}
R^{\text{(J)}}= \frac{\pi^{2}}{3}+\frac{4}{3\tau_{a}^{2}}
\end{eqnarray}
and
\begin{eqnarray}\label{apAeqRM}
R^{\text{(M)}}= -\frac{\pi^{2}}{6}+\frac{4}{3\tau_{a}^{2}},
\end{eqnarray}
where J=(D,N) for Dirichlet and Neumann, and M=(DN, ND) holds for mixed conditions, with the configurations DN and ND, respectively. 
%

%%%%%%%%%%%%%%%%%%%%%%%%%%%%%%%%
\subsubsection{Position dependent term}
%%%%%%%%%%%%%%%%%%%%%%%%%%%%%%%%
%
In the late time regime, that is, $\tau_a\gg 1$, the term $P^{\text{(i)}}_{x_{a}}$, Eq. \eqref{apAeq01a}, can be written in the form
\begin{eqnarray}\label{apAeq04}
P^{\text{(i)}}_{x_{a}}=-2\sum_{k=0}^{\infty}\left(\frac{2}{\tau_{a}}\right)^{2k}\sum_{n=-\infty}^{\infty}\frac{\delta_{n}^{\text{(i)}}}{(x_{a}-n)^{2-2k}},
\end{eqnarray}
where we have considered a series expansion for the denominator of $P^{\text{(i)}}_{x_{a}}$, and $\delta_{n}^{\text{(D)}}=-1$ and $\delta_{n}^{\text{(N)}}=1$ for Dirichlet and Neumann conditions, respectively. These coefficients differ by one sign and they are independent of the summation index. So from Eq. \eqref{apAeq04} we can write
\begin{eqnarray}\label{apAeq05}
P^{\text{(J)}}_{x_{a}}&=&-2\delta^{\text{(J)}}\sum_{k=0}^{\infty}\left(\frac{2}{\tau_{a}}\right)^{2k}\sum_{n=-\infty}^{\infty}\frac{1}{(x_{a}-n)^{2-2k}}\nonumber\\
&=&-2\delta^{\text{(J)}}\sum_{k=0}^{\infty}\left(\frac{2}{\tau_{a}}\right)^{2k} \left[             \frac{-1}{x_{a}^{2-2k}}+\sum_{j=\pm 1}\sum_{n=0}^{\infty}\frac{1}{(n+jx_{a})^{2-2k}}\right],
\end{eqnarray}
with $\delta^{\text{(J)}}=[\delta^{\text{(D)}},\delta^{\text{(N)}}]=[-1,+1]$. To achieve the second equality we have divided the initial summation in two parts and re-labeled the summation index of the negative interval. Next, we have written the two parts, with denominators of opposite signs, in a compact form by means of the $j$ summation.

By using the relation%, $5.1.3(1)$ de \cite[p. 652]{prudnikov1986integralsvol1},
\begin{eqnarray}\label{apAprop03}
\sum_{k=0}^{\infty}\frac{1}{(k+a)^{s}}=\zeta(s,a),\qquad\qquad \text{Re}(s)>1,
\end{eqnarray}
in Eq. \eqref{apAeq05} we obtain that
\begin{eqnarray}\label{ApAeq06}
P^{\text{(J)}}_{x_{a}}&=&-2\delta^{\text{(J)}}\left[ -\frac{x_{a}^{-2}}{[1-(2x_{a}/\tau_{a})^{2}]} + \left(\frac{2}{\tau_{a}}\right)^{2}\sum_{j=\pm 1}\sum_{m=-1}^{\infty}\left(\frac{2}{\tau_{a}}\right)^{2m}\zeta(-2m,jx_{a}) \right],
\end{eqnarray}
where we have performed the summation of the first term and re-labeled the sum index on the second term.

Finally, by making use of the Bernoulli polynomials
\begin{eqnarray}\label{apAprop04}
\zeta(-n,q) = -\frac{B_{n+1}(q)}{n+1},
\end{eqnarray}
where $n$ is a nonnegative integer and%, Eq. $9.531$ de \cite[p. 1037]{gradshtein2007}, e 
\begin{eqnarray}\label{apAprop05}
(-1)^{n}B_{n}(-x)=B_{n}(x)+nx^{n-1},
\end{eqnarray}
%Eq. $9.623(5)$ de \cite[p. 1042]{gradshtein2007}, 
after some algebraic work, we find for Eq. \eqref{ApAeq06} 
\begin{eqnarray}\label{apAeq07}
P^{\text{(J)}}_{x_{a}}=-\delta^{\text{(J)}}2\pi^{2}\csc^{2}(\pi x_{a}).
\end{eqnarray}

Now for mixed boundary conditions $\delta_{n}^{\text{(DN)}}=(-1)^{n+1}$ and $\delta_{n}^{\text{(ND)}}=(-1)^{n}$. From Eq. \eqref{apAeq04}, we have
\begin{eqnarray}\label{apAeq08}
P^{\text{(M)}}_{x_{a}}&=&2\delta^{\text{(M)}}\sum_{k=0}^{\infty}\left(\frac{2}{\tau_{a}}\right)^{2k}\sum_{n=-\infty}^{\infty}\frac{(-1)^{n}}{(x_{a}-n)^{2-2k}},\nonumber\\
&=&2\delta^{\text{(M)}}\sum_{k=0}^{\infty}\left(\frac{2}{\tau_{a}}\right)^{2k} \left[ -\frac{1}{x_{a}^{2-2k}}+\sum_{j=\pm 1}\sum_{n=0}^{\infty}\frac{(-1)^{n}}{(n+jx_{a})^{2-2k}}\right],
\end{eqnarray}
where $\delta^{\text{(M)}}=[\delta^{\text{(DN)}},\delta^{\text{(ND)}}]=[+1,-1]$. To establish the second equality we have performed a similar procedure to that used for $P^{\text{(J)}}$ (see text below Eq. \eqref{apAeq05}).

With the relation%, $5.1.3(5)$ de \cite[p. 653]{prudnikov1986integralsvol1},
\begin{eqnarray}\label{apAprop06}
\sum_{k=0}^{\infty}\frac{(-1)^{k}}{(k+a)^{s}}=2^{-s}\left[ \zeta\left(s,\frac{a}{2}\right)-\zeta\left(s,\frac{a+1}{2}\right) \right],\qquad\qquad \text{Re}(s)>0,
\end{eqnarray}
we can perform the summation on $n$ in Eq. \eqref{apAeq08} and, after a suitable index change, use Eq. \eqref{apAprop04} to write the resulting expression in terms of Bernoulli polynomials. Then, by considering Eqs. \eqref{apAeq07}, \eqref{apAprop06}, \eqref{apAprop04} and the identity %, $9.623(4)$ de \cite[p. 1042]{gradshtein2007}
\begin{eqnarray}\label{apAprop07}
B_{n}(1-x)=(-1)^{n}B_{n}(x),
\end{eqnarray}
we found
\begin{eqnarray}\label{apAeq09}
P_{x_{a}}^{\text{(M)}}= \delta^{\text{(M)}}2\pi^{2}\cot(\pi x_{a})\csc(\pi x_{a}).
\end{eqnarray}

For the function $Q^{\text{(i)}}_{x_{a}}$ we can write %an expression similar to the structures that we have seen so far, namely, 
\begin{eqnarray}\label{apAeq10}
Q^{\text{(i)}}_{x_{a}} = 2\tau_{a}\sum_{k=1}^{\infty}\frac{1}{(2k-1)}\left(\frac{2}{\tau_{a}}\right)^{2k-1}\sum_{n=-\infty}^{\infty}\frac{\delta_{n}^{\text{(i)}}}{(n-x_{a})^{4-2k}},
\end{eqnarray}
where we have used Eq. \eqref{apAprop02}. All mathematical manipulations are similar to those used so far, so in order to avoid repetitions we will be more succinct.

For Dirichlet and Neumann boundary conditions, we obtain
\begin{eqnarray}\label{apAeq11}
Q^{\text{(J)}}_{x_{a}} &=& \tau_{a}^{2}\delta^{\text{(J)}}\sum_{k=1}^{\infty}\frac{1}{(2k-1)}\left(\frac{2}{\tau_{a}}\right)^{2k}\left[- \frac{1}{x_{a}^{4-2k}} + \sum_{j=\pm 1}\sum_{n=0}^{\infty} \frac{1}{(n+jx_{a})^{4-2k}}\right],\nonumber\\
 &=&\delta^{\text{(J)}}4\pi^{2}\csc^{2}(\pi x_{a}).
\end{eqnarray}
Note that in the first equality we have used Eq. \eqref{apAprop03} to perform the sum in $n$ and Eq. \eqref{apAprop04} to express the solution in terms of the Bernoulli polynomials. Next, we have used the identity \eqref{apAprop05} to develop the resulting expression and achieve the result shown in the second equality.

In the case of mixed boundary conditions we have
\begin{eqnarray}\label{apAeq12}
Q^{\text{(M)}}_{x_{a}} &=&- \tau_{a}^{2}\delta^{\text{(J)}}\sum_{k=1}^{\infty}\frac{1}{(2k-1)}\left(\frac{2}{\tau_{a}}\right)^{2k}\left[ -\frac{1}{x_{a}^{4-2k}} + \sum_{j=\pm 1}\sum_{n=0}^{\infty} \frac{(-1)^{n}}{(n+jx_{a})^{4-2k}}\right],\nonumber\\
&=&-\delta^{\text{(M)}}4\pi^{2}\cot(\pi x_{a})\csc(\pi x_{a}).
\end{eqnarray}
Again, to establish the above result we have followed a similar procedure to the previous case, first we use Eq. \eqref{apAprop06} to perform the summation in the index $n$ and Eq. \eqref{apAprop04} to write the result in terms of Bernoulli polynomials. Then, we conclude the computation by observing the identities \eqref{apAprop05} and \eqref{apAprop07}.

In view of the results obtained for the position dependent functions, namely, Eqs. \eqref{apAeq07}, \eqref{apAeq09}, \eqref{apAeq11} and \eqref{apAeq12}, we find for the Dirichlet and Neumann boundary conditions 
\begin{eqnarray}\label{apAeqRxDN}
R_{x_{a}}^{\text{(J)}} = \delta^{\text{(J)}}2\pi^{2}\csc^{2}(\pi x_{a})
\end{eqnarray}
while for mixed boundary conditions we have
\begin{eqnarray}\label{apAeqRxM}
R_{x_{a}}^{\text{(M)}} = - \delta^{\text{(M)}}2\pi^{2}\cot(\pi x_{a})\csc(\pi x_{a}),
\end{eqnarray}
where $\delta^{\text{(J)}} = [\delta^{\text{(D)}},\delta^{\text{(N)}}]=[-1,+1]$ and $\delta^{\text{(M)}} = [\delta^{\text{(DN)}},\delta^{\text{(ND)}}]=[+1,-1]$. 
%
%

%%%%%%%%%%%%%%%%%%%%%%%%%%%%%%%%%%%%%%%%%%%%%
\subsection{Quasiperiodic condition}\label{AppAQPlatetime}
%%%%%%%%%%%%%%%%%%%%%%%%%%%%%%%%%%%%%%%%%%%%%
Similar to the approach introduced for the case of parallel planes, initially, we conveniently define the quantities
\begin{eqnarray}\label{apAqpeq00}
U(\beta,\tau_{a}):= \sum_{n=1}^{\infty}U(n,\beta,\tau_{a}) = S(\beta,\tau_{a})+T(\beta,\tau_{a}),
\end{eqnarray}
with
\begin{eqnarray}\label{apAqpeq00a}
S(\beta,\tau_{a}):= \sum_{n=1}^{\infty}S(n,\beta,\tau_{a})
\end{eqnarray}
and
\begin{eqnarray}\label{apAqpeq00b}
T(\beta,\tau_{a}):= \sum_{n=1}^{\infty}T(n,\beta,\tau_{a}).
\end{eqnarray}
The functions $U(n,\beta,\tau_{a})$, $S(n,\beta,\tau_{a})$, and $T(n,\beta,\tau_{a})$ shown above are defined in Eqs. \eqref{eq34}, \eqref{eq34a}, and \eqref{eq34b}, respectively.

In order to work out Eq. \eqref{apAqpeq00a} we observe that we can write
\begin{eqnarray}\label{apAqpeq01}
S(\beta,\tau_{a}) = -\frac{1}{(\tau_{a})^{2}}\sum_{m=-1}^{\infty}\frac{1}{(\tau_{a})^{2m}}\sum_{n=1}^{\infty}\frac{\cos(2\pi\beta n)}{n^{-2m}},
\end{eqnarray}
where we have considered a series expansion for the denominator and re-labeled the summation index. The first term of Eq. \eqref{apAqpeq01}, $m=-1$, can be solved using the relation
\begin{eqnarray}\label{apAqpprop00} %5.4.2(7), \cite[p. 726]{prudnikov1986integralsvol1}
\sum_{k=1}^{\infty}\frac{\cos(kx)}{k^{2n}} = \frac{(-1)^{n-1}(2\pi)^{2n}}{2(2n)!}B_{2n}\left(\frac{x}{2\pi}\right),
\end{eqnarray}
where $0\leq x\leq 2\pi$, $ n=1,2,\ldots$, and $B_{n}(z)$ are the Bernoulli polynomials of order $n$ in the variable $z$. The remained terms of the series in Eq. \eqref{apAqpeq01} can be computed by using the cosine series formula,
\begin{eqnarray}\label{apAqpprop01} % 1.411(3), \cite[p. 42]{gradshtein2007}
\cos(x)=\sum_{k=0}^{\infty}\frac{(-1)^{k}x^{2k}}{(2k)!},
\end{eqnarray}
and observing that $\zeta(-2n)=0$, where $n =1,2,\ldots $\;. So, we obtain
\begin{eqnarray}\label{apApqeq02}
S(\beta,\tau_{a}) = -\pi^{2}B_{2}(\beta) + \frac{1}{2\tau_{a}^{2}}.
\end{eqnarray}

In a similar way we have
\begin{eqnarray}\label{apApqeq03}
T(\beta,\tau_{a}) = \frac{2}{\tau_{a}^{2}}\sum_{m=-1}^{\infty} \frac{\tau_{a}^{-2m}}{(2m+3)} \sum_{n=1}^{\infty}\cos(2\pi\beta n)n^{2m},
\end{eqnarray}
which from Eqs. \eqref{apAqpprop00}, \eqref{apAqpprop01} and following the same arguments previously presented, give us
\begin{eqnarray}\label{apApqeq04}
T(\beta,\tau_{a}) = 2\pi^{2}B_{2}(\beta) - \frac{1}{3\tau_{a}^{2}}.
\end{eqnarray}

In view of the results \eqref{apApqeq02} and \eqref{apApqeq04}, from the Eq. \eqref{apAqpeq00}, we found
\begin{eqnarray}\label{apApqeq05}
U(\beta,\tau_{a}) = \pi^{2}B_{2}(\beta) + \frac{1}{6\tau_{a}^{2}}.
\end{eqnarray}
We stress that the above equation is an expression resulting from Eq. \eqref{apAqpeq00} for the late time regime, $\tau_a\gg 1$. We have numerically check that this is in fact the case.
%
%
%

%%%%%%%%%%%%%%%%%%%%%%%%%%%%%%%%%%%%%%%
\section{Short time regime}\label{AppB}
%%%%%%%%%%%%%%%%%%%%%%%%%%%%%%%%%%%%%%%
%
In this part, we shall analyze the expressions $R(n,\tau_{a})$ and $R(x_{a}-n,\tau_{a})$ for the short time regime, that is, $\tau_{a}\ll 1$. The methodology adopted is similar to that in Appendix \ref{AppA} and all mathematical relations used below can be found in Refs. \cite{prudnikov1986integralsvol1, prudnikov1986integrals, gradshtein2007}.  In addition, since the method used here is similar to the one used in the late time regime previously, we shall be more straightforward about the details, but we indicate the crucial steps when necessary.

\subsection{Dirichlet, Neumann and mixed boundary conditions}\label{AppBDNMshorttime}

\subsubsection{Position independent term}

First we write Eq. \eqref{apAeq00} in form
\begin{eqnarray}\label{apBeq00}
P^{\text{(i)}}=\frac{\tau_{a}^{2}}{2}\sum_{k=0}^{\infty}\left(\frac{\tau_{a}}{2}\right)^{2k}\sum_{n=1}^{\infty}\frac{\gamma_{n}^{\text{(i)}}}{n^{2k+4}}.
\end{eqnarray}
By noting that for Dirichlet and Neumann boundary conditions $\gamma_{n}^{\text{(D)}}=\gamma_{n}^{\text{(N)}}=1$, from Eqs. \eqref{apAprop00} and \eqref{apBeq00} we obtain
\begin{eqnarray}\label{apBeq01}
P^{\text{(J)}}=\frac{8}{\tau_{a}^{2}}\sum_{m=2}^{\infty}\left(\frac{\tau_{a}}{2}\right)^{2m}\zeta(2m),
\end{eqnarray}
where we have re-labeled the index summation. Now, by using%, $5.3.1(5)$ of \cite[p. 648]{prudnikov1986integrals}
\begin{eqnarray}\label{apBprop00}
\sum_{k=0}^{\infty}(\pm 1)^{k}t^{2k}\zeta(2k)=-\frac{\pi t}{2}\left\{
	\begin{matrix}
		\cot(\pi t)	&	\\
		\coth(\pi t) 
	\end{matrix}\right. , \qquad\qquad |t|<1,
\end{eqnarray}
we obtain
\begin{eqnarray}\label{apBeq02}
P^{\text{(J)}}= -\frac{\pi^{2}}{3}+\frac{4}{\tau_{a}^{2}}-\frac{2\pi}{\tau_{a}}\cot\left(\frac{\pi\tau_{a}}{2}\right),
\end{eqnarray}
with J=(D,N).

In the case of mixed boundary conditions, $\gamma_{n}^{\text{(DN)}}=\gamma_{n}^{\text{(ND)}}=(-1)^{n}$. Then, from Eq. \eqref{apBeq00} we have
\begin{eqnarray}\label{apBeq03}
P^{\text{(M)}}=\frac{\tau_{a}^{2}}{2}\sum_{k=0}^{\infty}\left(\frac{\tau_{a}}{2}\right)^{2k}\sum_{n=1}^{\infty}\frac{(-1)^{n}}{n^{2k+4}},
\end{eqnarray}
which by using Eq. \eqref{apAprop01} provides
\begin{eqnarray}\label{apBeq04}
P^{\text{(M)}}=-\frac{8}{\tau_{a}^{2}}\sum_{m=2}^{\infty}\left(\frac{\tau_{a}}{2}\right)^{2m}\zeta(2m)+\frac{16}{\tau_{a}^{2}}\sum_{m=2}^{\infty}\left(\frac{\tau_{a}}{4}\right)^{2m}\zeta(2m).
\end{eqnarray}
By considering Eq. \eqref{apBprop00} in the above expression we found
\begin{eqnarray}\label{apBeq05}
P^{\text{(M)}} = \frac{\pi^{2}}{6}+\frac{4}{\tau_{a}^{2}}-\frac{\pi}{\tau_{a}}\csc\left(\frac{\pi\tau_{a}}{4}\right)\sec\left(\frac{\pi\tau_{a}}{4}\right),
\end{eqnarray}
with M=(DN, ND).

Similarly, from Eq. \eqref{apAeq00b} we can write
\begin{eqnarray}\label{apBeq06}
Q^{\text{(i)}}=4\sum_{k=1}^{\infty}\frac{1}{(2k-1)}\left(\frac{\tau_{a}}{2}\right)^{2k}\sum_{n=1}^{\infty}\frac{\gamma_{n}^{(i)}}{n^{2k+2}},
\end{eqnarray}
which for Dirichlet and Neumann conditions provides 
\begin{eqnarray}\label{apBeq07}
Q^{\text{(J)}}=\frac{16}{\tau_{a}^{2}}\sum_{m=2}^{\infty}\frac{\zeta(2m)}{(2m-3)}\left(\frac{\tau_{a}}{2}\right)^{2m},
\end{eqnarray}
where we have made use of Eqs. \eqref{apAprop02} and \eqref{apAprop00}.

For mixed condition, from Eq. \eqref{apBeq06}, we have
\begin{eqnarray}\label{apBeq08}
Q^{\text{(M)}}=4\sum_{k=1}^{\infty}\frac{1}{(2k-1)}\left(\frac{\tau_{a}}{2}\right)^{2k}\sum_{n=1}^{\infty}\frac{(-1)^{n}}{n^{2k+2}},
\end{eqnarray}
which by using Eq. \eqref{apAprop06} gives us
\begin{eqnarray}\label{apBeq09}
Q^{\text{(M)}}=\frac{16}{\tau_{a}^{2}}\left[-\sum_{m=2}^{\infty}\frac{\zeta(2m)}{(2m-3)}\left(\frac{\tau_{a}}{2}\right)^{2m} + 2\sum_{m=2}^{\infty}\frac{\zeta(2m)}{(2m-3)}\left(\frac{\tau_{a}}{4}\right)^{2m}\right].
\end{eqnarray}
Since we have found useful expressions for $P^{(i)}$ and $Q^{(i)}$ we can easily obtain the corresponding functions $R^{(i)}$ in the short time regime. Regarding the series in the  functions $Q^{(i)}$, since we are working in the regime $\tau_{a}\ll 1$, is sufficient to consider the leading term in the power series of $\tau_{a}$. Thus, for Dirichlet and Neumann conditions, Eqs. \eqref{apBeq02} and \eqref{apBeq07}, we found
\begin{eqnarray}\label{apBeqRDN}
R^{\text{(J)}} &=& -\frac{\pi^2}{3}+\frac{4}{\tau_{a}^{2}}-\frac{2\pi}{\tau_{a}}\cot\left(\frac{\pi\tau_{a}}{2}\right)+\tau_{a}^{2}\zeta(4),\nonumber\\
&\simeq & \frac{3\tau_{a}^{2}}{2}\zeta(4).
\end{eqnarray}
On the other hand, for mixed boundary conditions, Eqs. \eqref{apBeq05} and \eqref{apBeq09}, we have
\begin{eqnarray}\label{apBeqRM}
R^{\text{(M)}} &=& \frac{\pi^2}{6}+\frac{4}{\tau_{a}^{2}}-\frac{\pi}{\tau_{a}}\csc\left(\frac{\pi\tau_{a}}{4}\right)\sec\left(\frac{\pi\tau_{a}}{4}\right)-\frac{7\tau_{a}^{2}}{8}\zeta(4),\nonumber\\
&\simeq & -\frac{21\tau_{a}^{2}}{16}\zeta(4).
\end{eqnarray}
In both cases, Eqs. \eqref{apBeqRDN} and \eqref{apBeqRM}, the expressions are only valid for $\tau_{a}\ll1$.
%
%

%%%%%%%%%%%%%%%%%%%%%%%%%%%%%%%%%%%%%%%%%%%%%
\subsubsection{Position dependent term}
%%%%%%%%%%%%%%%%%%%%%%%%%%%%%%%%%%%%%%%%%%%%%
%
From Eq. \eqref{apAeq01a}, firstly we write
\begin{eqnarray}\label{apBeq10}
P_{x_{a}}^{\text{(i)}} = \frac{\tau_{a}^{2}}{2}\sum_{k=0}^{\infty}\left(\frac{\tau_{a}}{2}\right)^{2k}\sum_{n=-\infty}^{\infty}\frac{\delta_{n}^{\text{(i)}}}{(n-x_{a})^{2k+4}}.
\end{eqnarray}
By considering Dirichlet and Neumann boundary conditions, namely, $\delta_{n}^{\text{(D)}}=-1$ and $\delta_{n}^{\text{(N)}}=1$, respectively, we can divide the summation in Eq. \eqref{apBeq10} in two parts and relabel the sum index to the negative range. This gives 
\begin{eqnarray}\label{apBeq11}
P_{x_{a}}^{\text{(J)}}= \delta^{\text{(J)}}\left[- \frac{\tau_{a}^{2}}{2x_{a}^{4}}\sum_{k=0}^{\infty}\left(\frac{\tau_{a}}{2x_{a}}\right)^{2k}  + \frac{8}{\tau_{a}^{2}}\sum_{j=\pm 1}\sum_{m=2}^{\infty}\left(\frac{\tau_{a}}{2}\right)^{2m}\zeta(2m,jx_{a}) \right],
\end{eqnarray}
where $\delta^{\text{(J)}}=[\delta^{\text{(D)}},\delta^{\text{(N)}}]=[-1,+1]$ and we have used Eq. \eqref{apAprop03} to perform the summation in the $n$ index.

The first term in the r.h.s. of Eq. \eqref{apBeq11} can easily be computed and the second term can also be solved using the integral representation %$9.511$ of \cite[p. 1036]{gradshtein2007},
\begin{eqnarray}\label{eqBprop01}
\zeta(z,q) = \frac{1}{\Gamma(z)}\int_{0}^{\infty}\frac{t^{z-1}e^{-qt}}{1-e^{-t}}dt.
\end{eqnarray}
Therefore, we obtain
\begin{eqnarray}\label{apBeq12}
P_{x_{a}}^{\text{(J)}}=- \delta^{\text{(J)}}\left\{2\pi^{2}\csc^{2}(\pi x_{a}) + \frac{2\pi}{\tau_{a}}\left[ \cot\left[\frac{\pi(\tau_{a}-2x_{a})}{2}\right] + \cot\left[\frac{\pi(\tau_{a}+2x_{a})}{2}\right]\right]   \right\}.
\end{eqnarray}

Now, from Eq. \eqref{apBeq10}, the mixed condition case provides
\begin{eqnarray}\label{apBeq13}
P_{x_{a}}^{\text{(M)}} = -\delta^{\text{(M)}} \frac{\tau_{a}^{2}}{2}\sum_{k=0}^{\infty}\left(\frac{\tau_{a}}{2}\right)^{2k}\sum_{n=-\infty}^{\infty}\frac{(-1)^{n}}{(n-x_{a})^{2k+4}}.
\end{eqnarray}
Consequently, by using Eq. \eqref{apAprop06} we found
\begin{eqnarray}\label{apBeq14}
P_{x_{a}}^{\text{(M)}}= \delta^{\text{(M)}}\left\{ \frac{\tau_{a}^{2}}{2x_{a}^{4}}\sum_{k=0}^{\infty}\left(\frac{\tau_{a}}{2x_{a}}\right)^{2k}  - \frac{8}{\tau_{a}^{2}}\sum_{j=\pm 1}\sum_{m=2}^{\infty}\left(\frac{\tau_{a}}{4}\right)^{2m}\left[\zeta\left(2m,\frac{jx_{a}}{2}\right) - \zeta\left(2m,\frac{1+ jx_{a}}{2}\right)\right]\right\}.
\end{eqnarray}
Furthermore, by making use of Eq. \eqref{eqBprop01} in the above expression, we can first perform the sums and then the integrals. Thus, after some algebraic work we obtain
\begin{eqnarray}\label{apBeq15}
P_{x_{a}}^{\text{(M)}}=\delta^{\text{(M)}}\left\{2\pi^{2}\cot(\pi x_{a})\csc(\pi x_{a}) + \frac{2\pi}{\tau_{a}}\left[ \csc\left[\frac{\pi(2x_{a} + \tau_{a})}{2}\right] - \csc\left[\frac{\pi(2x_{a}-\tau_{a})}{2}\right]\right] \right\},
\end{eqnarray}
where $\delta^{\text{(M)}} = [\delta^{\text{(DN)}},\delta^{\text{(ND)}}]=[+1,-1]$.

For the function $Q^{(i)}_{x_{a}}$, from Eq. \eqref{apAeq01b}, we have
\begin{eqnarray}\label{apBeq16}
Q^{\text{(i)}}_{x_{a}}=4\sum_{k=1}^{\infty}\frac{1}{(2k-1)}\left(\frac{\tau_{a}}{2}\right)^{2k}\sum_{n=-\infty}^{\infty}\frac{\delta_{n}^{\text{(i)}}}{(n-x_{a})^{2k+2}}.
\end{eqnarray}
In the case of Dirichlet and Neumann conditions
\begin{eqnarray}\label{apBeq17}
Q^{\text{(J)}}_{x_{a}}=\delta^{\text{(J)}}\frac{16}{\tau_{a}^{2}}\sum_{m=2}^{\infty}\frac{1}{(2m-3)}\left(\frac{\tau_{a}}{2}\right)^{2m}\left[ -\frac{1}{x_{a}^{2m}}  + \sum_{j=\pm 1}\sum_{n=0}^{\infty}  \frac{1}{(n+jx_{a})^{2m}}\right],
\end{eqnarray}
which from Eq. \eqref{apAprop03} gives us
\begin{eqnarray}\label{apBeq18}
Q^{\text{(J)}}_{x_{a}}=\delta^{\text{(J)}}\left[-\frac{\tau_{a}}{2x_{a}^{3}}\ln\left(\frac{2x_{a}+\tau_{a}}{2x_{a}-\tau_{a}}\right)^{2} +  \frac{16}{\tau_{a}^{2}} \sum_{j=\pm 1}\sum_{m=2}^{\infty}\frac{\zeta(2m,jx_{a})}{(2m-3)}\left(\frac{\tau_{a}}{2}\right)^{2m}  \right].
\end{eqnarray}

The mixed condition case has an expression very similar to the one in Eq. \eqref{apBeq17}, namely,
\begin{eqnarray}\label{apBeq19}
Q^{\text{(M)}}_{x_{a}}=-\delta^{\text{(M)}}\frac{16}{\tau_{a}^{2}}\sum_{m=2}^{\infty}\frac{1}{(2m-3)}\left(\frac{\tau_{a}}{2}\right)^{2m}\left[ -\frac{1}{x_{a}^{2m}}  + \sum_{j=\pm 1}\sum_{n=0}^{\infty}  \frac{(-1)^{n}}{(n+jx_{a})^{2m}}\right],
\end{eqnarray}
which by mean of Eq. \eqref{apAprop06} provides
\begin{eqnarray}\label{apBeq20}
Q^{\text{(M)}}_{x_{a}}=\delta^{\text{(M)}}\left\{\frac{\tau_{a}}{2x_{a}^{3}}\ln\left(\frac{2x_{a}+\tau_{a}}{2x_{a}-\tau_{a}}\right)^{2} -  \frac{16}{\tau_{a}^{2}} \sum_{j=\pm 1}\sum_{m=2}^{\infty}\frac{1}{(2m-3)} \left(\frac{\tau_{a}}{4}\right)^{2m}\left[ \zeta\left(2m,\frac{jx_{a}}{2}\right)- \zeta\left(2m,\frac{1+jx_{a}}{2}\right) \right]   \right\}.\nonumber\\
\end{eqnarray}

Finally, with the useful expressions for $P_{x_{a}}^{(i)}$ and $Q_{x_{a}}^{(i)}$ we can use Eq. \eqref{apAeq01} to construct the functions $R_{x_{a}}^{(i)}$. Thereby, for Dirichlet and Neumann conditions, from Eqs. \eqref{apBeq12} and \eqref{apBeq18}, we obtain 
\begin{eqnarray}\label{apBeq21}
R^{\text{(J)}}_{x_{a}}&=& \delta^{\text{(J)}}\left\{-2\pi^{2}\csc^{2}(\pi x_{a}) - \frac{2\pi}{\tau_{a}}\left[
\cot\left[\frac{\pi(\tau_{a}-2x_{a})}{2}\right]+\cot\left[\frac{\pi(\tau_{a}+2x_{a})}{2}\right] \right]\right.\nonumber\\
&-&\left. \frac{\tau_{a}}{2x_{a}^{3}}\ln\left(\frac{2x_{a}+\tau_{a}}{2x_{a}-\tau_{a}}\right)^{2} + \tau_{a}^{2}[\zeta(4,x_{a})+\zeta(4,-x_{a})]\right\},\nonumber\\
&\simeq & \delta^{\text{(J)}}\left\{ \frac{\tau_{a}^{2}\pi^{4}}{2}[2+\cos(2\pi x_{a})]\csc^{4}(\pi x_{a})   \right\},
\end{eqnarray}
whereas for mixed boundary conditions, from Eqs. \eqref{apBeq15} and \eqref{apBeq20}, we have
\begin{eqnarray}\label{apBeq22}
R^{\text{(M)}}_{x_{a}}&=& \delta^{\text{(M)}}\left\{ 2\pi^{2}\cot(\pi x_{a})\csc(\pi x_{a}) + \frac{2\pi}{\tau_{a}}\left[\csc\left[\frac{\pi(2x_{a}+\tau_{a})}{2}\right]+\csc\left[\frac{\pi(2x_{a}-\tau_{a})}{2}\right] \right]\right.\nonumber\\
&+&\left. \frac{\tau_{a}}{2x_{a}^{3}}\ln\left(\frac{2x_{a}+\tau_{a}}{2x_{a}-\tau_{a}}\right)^{2} - \frac{\tau_{a}^{2}}{16}\left[\zeta\left(4,\frac{x_{a}}{2}\right)+\zeta\left(4,\frac{-x_{a}}{2}\right)-\zeta\left(4,\frac{1+x_{a}}{2}\right)-\zeta\left(4,\frac{1-x_{a}}{2}\right)\right]\right\},\nonumber\\
&\simeq & \delta^{\text{(M)}}\left\{ \frac{-\tau_{a}^{2}\pi^{4}}{8}[11+\cos(2\pi x_{a})]\cot(\pi x_{a})\csc^{3}(\pi x_{a})   \right\},
\end{eqnarray}
with $\delta^{\text{(J)}} = [\delta^{\text{(D)}},\delta^{\text{(N)}}]=[-1,+1]$ and $\delta^{\text{(M)}} = [\delta^{\text{(DN)}},\delta^{\text{(ND)}}]=[+1,-1]$. It is important to point out that, since we are working in the regime $\tau_{a}\ll 1$, we have only considered the leading term for the series of the functions $Q_{x_{a}}^{(i)}$. 

\subsection{Quasiperiodic condition}\label{AppBQPshorttime}

The function $S(\beta,\tau_{a})$, Eq. \eqref{apAqpeq00a}, can be worked out by observing that 
\begin{eqnarray}\label{apBqpeq23}
S(\beta,\tau_{a}) =\sum_{n=1}^{\infty}\frac{\tau_{a}^{2}\cos(2\pi\beta n)}{n^{2}(n^{2}-\tau_{a}^{2})} = \frac{1}{\tau_{a}^{2}}\sum_{m=2}^{\infty}(\tau_{a})^{2m}\sum_{n=1}^{\infty}\frac{\cos(2\pi\beta n)}{n^{2m}},
\end{eqnarray}
where we have first considered a series expansion for the denominator and re-labeled the index sum. So, by using the relation in Eq. \eqref{apAqpprop00}, we find that
\begin{eqnarray}\label{apBqpeq24}
S(\beta,\tau_{a}) = -\frac{1}{2\tau_{a}^{2}}\sum_{m=2}^{\infty}\frac{(-1)^{m}(2\pi\tau_{a})^{2m}}{(2m)!} B_{2m}(\beta),
\end{eqnarray}
where $B_{2m}(\beta)$ are the Bernoulli polynomials of order $2m$ on the variable $\beta$.

Similarly, for the functions $T(\beta,\tau_{a})$, Eq. \eqref{apAqpeq00b}, using the relation \eqref{apAprop02} and redefining the sum index, we found
\begin{eqnarray}\label{apBqpeq25}
 T(\beta,\tau_{a}) &=& \frac{2}{\tau_{a}^{2}}\sum_{m=2}^{\infty}\frac{\tau_{a}^{2m}}{(2m-3)}\sum_{k=1}^{\infty}\frac{\cos(2\pi\beta n)}{n^{2m}},
\end{eqnarray}
which by using Eq. \eqref{apAqpprop00} provides
\begin{eqnarray}\label{apBqpeq26}
T(\beta,\tau_{a}) &=&- \frac{1}{\tau_{a}^{2}}\sum_{m=2}^{\infty}\frac{(-1)^{m}(2\pi\tau_{a})^{2m}}{(2m-3)(2m)!}B_{2m}(\beta).
\end{eqnarray}

Now, from Eqs. \eqref{apAqpeq00}, \eqref{apBqpeq24} and \eqref{apBqpeq26} we finally find the following approximation:
\begin{eqnarray}\label{apBqpeq27}
U(\beta,\tau_{a}) &=&- \frac{1}{2\tau_{a}^{2}}\sum_{m=2}^{\infty}\frac{(2m-1)(-1)^{m}(2\pi\tau_{a})^{2m}}{(2m-3)(2m)!}B_{2m}(\beta),\nonumber\\
&\simeq & - \pi^{4}\tau_{a}^{2}B_{4}(\beta),
\end{eqnarray}
where we have only considered the leading term of the series, since $\tau_{a}\ll 1$. Once again, the expressions obtained here are only valid for the short time regime, something that we have numerically checked.

%\end{appendices}

%\bibliographystyle{JHEP}
%\bibliography{ref.bib}

\end{document}